\documentclass{emulateapj}
\usepackage{graphicx,amsmath,color,natbib,flmanyeq}

\newcommand{\boldx}{\mathbf x}
\newcommand{\boldk}{\mathbf k}
\newcommand{\boldq}{\mathbf q}
\newcommand{\boldPsi}{\boldsymbol{\Psi}}
\newcommand{\deltahat}{\hat \delta}
\newcommand{\Omegam}{\Omega_{\rm m}}
\newcommand{\Omegal}{\Omega_{\Lambda}}
\newcommand{\Omegak}{\Omega_{\rm k}}
\newcommand{\mpch}{\ h^{-1} \rm{Mpc}}
\newcommand{\beq}{\begin{equation}}
\newcommand{\eeq}{\end{equation}}
\newcommand{\bea}{\begin{eqnarray}}
\newcommand{\eea}{\end{eqnarray}}
\newcommand{\dirac}{\delta_{\rm D}}
\newcommand{\shah}{\rm{I\!I\!I}}
\newcommand{\dbar}{\bar D}
\newcommand{\abox}{a_{\rm box}}
\newcommand{\auni}{a_{\rm uni}}
\newcommand{\hobox}{H_{0,{\rm box}}}
\newcommand{\houni}{H_{0,{\rm uni}}}
\newcommand{\ombox}{\Omega_{\rm m,box}}
\newcommand{\omuni}{\Omega_{\rm m,uni}}
\newcommand{\olbox}{\Omega_{\Lambda,{\rm box}}}
\newcommand{\oluni}{\Omega_{\Lambda,{\rm uni}}}
\newcommand{\Gyr}{\ {\rm Gyr}}

\begin{document}

\title{Initial conditions to cosmological $N$-body simulations, or 
	how to run an ensemble of simulations}
\author{Edwin~Sirko}
\email{esirko@astro.princeton.edu}
\affil{Princeton University Observatory, Princeton, NJ 08544--1001, USA}

\begin{abstract}
The conventional method of generating initial conditions for cosmological $N$-body simulations introduces a significant error in the real-space statistical properties of the particles.  More specifically, the finite box size leads to a significant underestimate of $\sigma_8$, the correlation function, and nonlinear effects.  I implement a method of generating initial conditions for $N$-body simulations that accurately models the real-space statistical properties, such as the mass variance in spheres and the correlation function.  The method requires running ensembles of simulations because the power in the DC mode is no longer assumed to be zero.  For moderately sized boxes, I demonstrate that the new method corrects the underestimate in the mass variance in spheres and the shape of the correlation function.  I also argue that subtracting Poisson noise from the power spectrum is a dangerous practice.  Code to generate initial conditions to second order in Lagrangian perturbation theory with the new method is available at 
\verb+http://www.astro.princeton.edu/~esirko/ic+.
\end{abstract}


\section{Introduction}
Cosmological $N$-body simulations are an important tool in understanding nonlinear structure formation \citep{bertschinger_1998}.  For a consistent framework, this paper focuses on the collisionless variety, where each particle represents a sampling of the underlying dark matter collissionless fluid and baryon physics is assumed to be negligible on large enough scales.  However, the main results of this paper can be applied to any type of numerical simulation, such as hydrodynamic simulations of galaxy formation and Ly$\alpha$ clouds, in which the mean density is representative of a cosmologically relevant part of the universe.  As with any integrator, the generation of initial conditions is as important as the integration itself.  However, there seems to be a relative dearth of information on initial conditions to $N$-body simulations in the literature, besides the standard ``we use the Zeldovich approximation'' tagline.  This paper is an attempt to fill that gap.

The approach for the present work was motivated by \citet{pen_1997}.  Pen pointed out that in order to accurately model certain statistics --- those most directly connected to analytic models of nonlinear structure formation --- one should ensure that the real-space statistical properties (e.g., the correlation function) of the initial conditions are accurately modeled, instead of the $k$-space statistical properties (e.g., the power spectrum) as is usually done.  In this paper, I denote initial conditions generated by the former method $\xi$-sampled IC's (because they accurately model the correlation function), and by the latter, $P$-sampled IC's (because they accurately model the power spectrum).  In the limit that the box size $L \to \infty$, the two methods converge to each other, but in general, it is impossible to accurately model the correlation function and power spectrum simultaneously in a finite box. 
\citet{bertschinger_2001} released a multiscale initial conditions generator called \textsc{grafic2}.  While building upon ideas from \citet{pen_1997}, \textsc{grafic2} does not actually incorporate Pen's crucial suggestion of using $\xi$-sampled IC's.  Bertschinger claimed that using $\xi$-sampled IC's would be inconsistent with $N$-body codes that use periodic boundary conditions, but there is actually no such inconsistency because $P$- and $\xi$-sampled IC's share the properties of periodic boundary conditions and discrete wavenumbers.

Historically, simulations were run until the mass variance in spheres $\sigma_8$ reached a certain value, and that epoch was called ``today.''  But with the advent of ``precision cosmology'' from accurate CMB measurements, $\sigma_8$ is no longer a fudge factor; it is a well-defined cosmological parameter that should be input into the initial conditions as a starting point to simulations.  Speaking broadly, the advent of precision cosmology requires simulations to be accurate at the percent level.  The motivation for this paper, therefore, is to get $N$-body simulations accurate enough for precision cosmology.

The discrepancy between $P$- and $\xi$-sampled IC's is closely related to the problem of finite box size.  By studying the fraction of mass in halos, \citet{bagla_ray_2004} have recently argued that box sizes should be much larger than those typically used.  But their analysis is for $P$-sampled IC's.  While I do not use the statistic of halo mass fraction in the present study, the analysis of $\sigma_8$ in \S\ref{s:sigma8} below implies that using $\xi$-sampled IC's solves the problem of underestimated real-space statistics in finite boxes ``to first order.''

\citet{joyce_levesque_marcos_2004} have proposed a new method of generating initial conditions by modifying the potential of a one-component plasma and integrating trajectories of particles (similar to generation of glass initial conditions).  They claim that this method will accurately model the correlation function and power spectrum simultaneously.  A clarification of this apparent contradiction is in order.  Their paper is an attempt to solve an orthogonal problem, that of ignoring finite box size effects but focusing on nonzero $\Delta x = L/N^{1/3}$ effects. 
In fact, particle discreteness has two adverse consequences. 
First, nearest-neighbor effects are dynamically important \citep{baertschiger_etal_2002}.  This consequence of particle discreteness seems impossible to alleviate without increasing the mass resolution of a given simulation.
Second, the correlation function of a perturbed lattice of particles is highly oscillatory and drastically misrepresents the true correlation function of the underlying continuous fluid \citep{baertschiger_syloslabini_2002,knebe_dominguez_2003,joyce_marcos_2004}.  The present paper ignores particle discreteness effects, instead focusing on the problem of finite box size $L$.
However, as seen in \S\ref{s:xi} below, the oscillations in the correlation function for a lattice are erased cleanly as the simulation evolves and the underlying lattice disappears.  It is reasonable to suppose that the effects of the oscillatory correlation function (which \citet{joyce_levesque_marcos_2004} attempt to solve) are not as important as the effects of a systematically underestimated correlation function (which the present paper attempts to solve).

As we will see, $\xi$-sampled IC's introduce a nonzero variance of the DC mode of a simulation.  This implies that running an \emph{ensemble} of simulations will be necessary, where each realization's DC mode is drawn from an appropriate distribution.  Running an ensemble is a good idea anyway, because of the small number of independent modes ($\sim 3$) with wavelength $\sim L$.  
To combat this latter effect of sample variance, 
\citet{knebe_dominguez_2003}, \citet{baugh_efstathiou_1994}, 
and \citet{baugh_gaztanaga_efstathiou_1995} 
ran ensembles of 10 simulations but used $P$-sampled IC's, ignoring the DC mode.  (The latter authors compensated for the lack of DC mode by using the technique of \citet{couchman_carlberg_1992} of extrapolating the power spectrum at small $k$ by linear theory.)  In contrast, various authors have proposed methods of extracting an ``ensemble's worth'' of information out of a single realization.  For example, a realization, being periodic, can be replicated and longer-wavelength modes linearly added \citep{tormen_bertschinger_1996,cole_1997}.  Similarly, \citet{couchman_carlberg_1992} linked nonlinear evolution at high $k$ with linear theory predictions at small $k$ in their simulations.  While these methods are useful in their own right, in this paper I will assume that computational resources are available to run an ensemble of simulations.  \citet{frenk_etal_1988} ran an ensemble of simulations very similar in spirit to the methods outlined in this paper, using $P$-sampled IC's for $\boldk \ne 0$, but taking into account the DC mode.  The present paper extends their method in several ways: the DC mode of each realization should be drawn from a Gaussian distribution, and all modes of the power spectrum, not just the DC mode, should be convolved with the window function for $\xi$-sampled IC's.

The next subsection, \S\ref{s:preliminaries}, is an exposition to some of the terminology and definitions used in this paper.  I outline the conventional method for generating initial conditions with the Zeldovich approximation in \S\ref{s:conventional_ic}.  \S\ref{s:sampling_pk} discusses the problem with the conventional method, and \S\ref{s:convolving_pk} discusses the solution of using $\xi$-sampled IC's.  \S\ref{s:dc_mode} considers the DC mode, an important complication arising from $\xi$-sampled IC's.  \S\ref{s:ensembles} presents numerical simulations designed to test $\xi$-sampled IC's.  \S\S\ref{s:sigma8}--\ref{s:pk} analyze certain statistics of the particle distributions: the mass variance in $8 \mpch$ spheres, the correlation function, and power spectrum, respectively.  A comment on subtracting the Poisson noise contribution follows in \S\ref{s:poisson}.  \S\ref{s:summary} is the summary.

\subsection{Definitions, notation, preliminaries, etc.}\label{s:preliminaries}

As is usual, we will assume that the cosmological simulation tracks the
evolution of a periodic comoving cubical box with side length $L$, so that the
simulated universe is at least homogenous out to infinity.  The volume $V=L^3$.
There are $N$ collisionless particles in the box, representing a sampling of
the dark matter field.  The early cosmological simulations of \citet{defw_1985}
used $N = 32^3$, and present-day state-of-the-art simulations by the Virgo consortium have used $2160^3$ particles \citep{springel_etal_2005}.

The density field $\rho$, whether continuous or discrete, becomes $\delta =
\rho/\rho_0 -1$ in dimensionless form, where $\rho_0$ is the average density.
If the density field consists of point particles, $\delta(\boldx) = \frac{V}{N}
\sum \dirac(\boldx - \boldx_{\rm p}) -1$ where $\dirac$ is the Dirac delta
function and $\boldx_{\rm p}$ is the location of each particle.  
The Fourier transform equations are (see \citet{he_1988}, Appendix A)
\begin{manyeqns}
\label{e:fourier}
\hat \delta(\boldk) &=& \int \delta(\boldx) e^{-i \boldk \cdot \boldx} 
	d^3\boldx \label{e:fourier1} \\
&=& \frac{V}{N} \sum e^{-i \boldk \cdot \boldx_{\rm p}} \qquad 
	(\boldk \ne 0) \nonumber \\
\delta(\boldx) &=& \frac{1}{V} \sum \hat \delta(\boldk) 
	e^{i \boldk \cdot \boldx} \label{e:fourier2}
\end{manyeqns}
Note equation~\ref{e:fourier2} can be derived by approximating the Fourier
transform integral $\delta(\boldx) = \frac{1}{(2\pi)^3} \int \hat
\delta(\boldk) e^{i \boldk \cdot \boldx} d^3\boldk$.  $N$-body codes follow the
trajectories of particles in a Lagrangian framework.  Despite the
discretization of matter into particles, the density field is continuous.  But
it occupies a finite volume, so the integral in equation~\ref{e:fourier1} is
over the box volume, the sum in equation~\ref{e:fourier2} is an infinite
series, and $\boldk$ is forced to have values $\in (i,j,k)\Delta k$ where
$\Delta k = 2\pi/L$ and $(i,j,k)$ are integers $\in (-\infty,\infty)$. For
comparison, a Eulerian discretization would turn the integral in
equation~\ref{e:fourier1} into a sum, and the sum in equation~\ref{e:fourier2}
would be Nyquist limited.  The interpretation of the density field given by
equation~\ref{e:fourier} is ``literal,'' in that the particles are delta
functions, and it is the convention I will use throughout this paper, but it is
worth keeping in mind two related issues.  First, $N$-body codes use force
softening to ensure that the evolution is collisionless. Force softening
imparts an effective particle shape, so this can be modeled as convolving the
``literal'' density field with the particle shape.  Thus the Fourier series
(equation~\ref{e:fourier2}) will effectively be limited at large wavenumber,
but not Nyquist limited.  Force softening, being a modification to the inverse-square force law, is often modeled as an effective particle shape interacting with a \emph{point} mass \citep{athanassoula_etal_2000, hernquist_katz_1989}.  However, in order to explore the effects of force softening on the power spectrum, an effective particle shape must be found that would be consistent with the modified force law \emph{if two of these nonzero-sized particles were interacting}.  The issue of force softening's effect on the power spectrum will not be considered further in this paper on the grounds that it is unlikely to shed much light on the regions of interest of the power spectrum.
Second, discretized $N$-body fields
(for cosmological collisionless simulations) are always interpreted as sampling
the underlying density field, which makes information about the power spectrum at wavelengths less than the typical particle separation meaningless.  However, it remains unclear if there is any ``best'' way to make
the conversion from a particle field to a continuous one, besides smoothing
with a large enough filter.  In any case, highly nonlinear effects at very high
$k$ are expected not to have much effect on the evolution of the modes of
interest in a simulation \citep[e.g.,][\S 28]{peebles_1980}.

The power spectrum of the finite-size box is $P(\boldk) = \frac{1}{V} \langle
|\hat\delta(\boldk)|^2 \rangle$ where the brackets indicate an ensemble
average.  Since the universe is isotropic, $P(k) = P(\boldk)$.  I define the
isotropic power spectrum of a specific realization of a set of particles in a
box with the same formula except with the average being over $|\boldk| = k$,
although there will be sample variance because of the finite number of modes.
Note that the power spectrum has units of volume, because it is a power
spectral density in $k$-space.  \citet{bertschinger_1992} has emphasized
that this convention should be used.

\section{The conventional method for generating initial conditions}\label{s:conventional_ic}

The goal of an initial conditions generator for a cosmological $N$-body
simulation is to faithfully reproduce the statistical properties of the density
field in the early universe with a finite number of point particles.  The
interpretation of a finite set of point particles as an underlying continuous
density field is somewhat ambiguous; hence, there are many ways to make this
conversion and it is not clear if there is any ``correct'' answer 
\citep[e.g.,][]{hernquist_katz_1989,bertschinger_gelb_1991,
weinberg_etal_1997,pelupessy_etal_2003,ascasibar_binney_2005}.
We have thus come across the first
thorny issue of initial conditions generation, before even discussing
cosmology: how does one convert a uniform density field into a set of discrete
particles?  The resultant particle distribution can appropriately be called
``pre-initial'' \citep[e.g.,][]{joyce_marcos_2004}.  It is simplest to use a
lattice of particles, where the position of each particle is given by $[(i,j,k)
+0.5] L/N^{1/3}$; $(i,j,k)$ are integer indices $\in [0, N^{1/3})$.  One
could also use glass pre-initial conditions, obtained by evolving randomly
placed particles with the sign of gravity reversed (with damping to help
convergence) \citep{white_1994, baugh_gaztanaga_efstathiou_1995}.
Both methods produce a particle distribution that is in
unstable equilibrium with zero gravitational force at the location of each
particle.  In contrast, a Poisson distribution would immediately evolve into
nonlinear structures, even though all three methods should recover the
continuum case in the limit of large $N$.  It seems better to use lattice or
glass pre-initial conditions because they contain no gravitational forces, but
it is not obvious that they solve all the problems of the Poisson pre-initial
distribution \citep{gabrielli_etal_2003}.

Another method is to modify the potential of a one-component plasma, which will
produce a particle distribution with the desired statistical properties of a
cosmological simulation without first generating pre-initial conditions
\citep{gabrielli_etal_2003, joyce_levesque_marcos_2004}.  So far this method is too
computationally expensive to be practical for large $N$-body simulations.

For simplicity I use a lattice for pre-initial conditions.  A lattice is just
an infinite series of delta functions at equal spacing, notated as $\shah$, and
pronounced ``shah'' \citep{bracewell_2000}.  Generally, like the Gaussian, the $\shah$ function is its own Fourier transform.  Particularly, the Fourier transform of our
dimensionless density particle lattice of spacing $\Delta x = L/N^{1/3}$ is
just $V$ for $\boldk = (i,j,k) 2 k_{\rm Ny}$ where $(i,j,k) \in$ all integers
except $(0,0,0)$, and zero otherwise, where the ``natural'' Nyquist 
frequency $k_{\rm Ny} = \pi/\Delta x$.
Thus $N$-body simulations must strike a compromise between
having a large enough box size $L$ while ensuring that all the modes of
interest are sub-Nyquist, so that the lattice pre-initial conditions contribute
no power in this regime.

Given a pre-initial particle distribution representing a completely homogeneous
universe, the next task is to perturb the particles to produce a density field
that reproduces the cosmologically relevant expected statistical properties.
I will go into some detail here because it will provide the framework for the algorithm
of \S\ref{s:dc_mode}, and also because most of the available literature on initial conditions is either quite abbreviated or, in some instances, incorrect.

In the conventional method, the power spectrum of the matter fluctuations in
the universe, determined from linear Boltzmann integrators such as 
cmbfast \citep{seljak_zaldarriaga_1996} 
or the earlier fit by \citet{bbks_1986}, 
is sampled on exactly those values of $\boldk$ which
are the abscissae for the finite-volume power spectrum.  
The power spectrum is usually calculated in linear theory and 
extrapolated to the present epoch.  Assuming a $\Lambda$CDM cosmology, in first order Eulerian perturbation theory all modes evolve independently, so the power spectrum can be scaled back to the initial epoch for the cosmological simulation via the growth function
\beq
D(a) = \frac{5 \Omegam H_0^2}{2} \frac{\dot a}{a} \int_0^a \frac{d a'} {\dot a'^3}
\label{e:growth_fxn}
\eeq
\citep[equation 7.77]{dodelson_2003} where dots denote derivatives with respect to proper time.  The time derivative of the growth function is
\beq
\dot{D}(a) =  \frac{\Omega_{\rm m} H_0^2}{2 \dot a a} \left[ 5 - \frac{3D}{a} - \frac{2 \Omega_{\rm k} D}{\Omega_{\rm m}} \right]. \label{e:growth_fxn_dot}
\eeq
We will mostly use the normalized growth function $\dbar(a) = \frac{D(a)}{D(1)}$.  The Hubble parameter at a given epoch is 
\beq
\frac{\dot a}{a} = H_0 \sqrt{ \Omegam/a^3 + \Omegal + \Omegak/a^2 }
\label{e:hubble}
\eeq
where $H_0$, $\Omegam$, and $\Omegal$ are parametrizations of the cosmology and $\Omegak = 1 - \Omegam - \Omegal$, and radiation is ignored here.

The power spectrum,
extrapolated to the present epoch in linear perturbation theory via
the growth equation,
is normalized to a given value of $\sigma_8$ such that
\bea
\sigma_8^2 &=& \frac{1}{(2\pi)^3} \int P(\boldk) \hat W^2(\boldk) d^3\boldk \nonumber \\
	&=& \frac{1}{2\pi^2} \int P(k) \hat W^2(k) k^2 dk
\eea
where $W(\boldx)$ is a top-hat spherical window function of radius $R = 8 \mpch$, normalized to integrate to $1$, whose Fourier transform is
\beq \label{e:wk}
\hat W(\boldk) = \hat W(k) = \frac{3}{(kR)^3}(\sin{kR} - kR\cos{kR}).
\eeq
For a Gaussian random field, a single realization should populate a given mode $\hat \delta(\boldk)$ with a value whose phase is random, and whose magnitude is drawn from a Rayleigh distribution:\footnote{Consider temporarily the limit $L \to \infty$.  The spacing of modes becomes infinitesimally small and the actual distribution used to populate magnitudes doesn't matter because the central limit theorem ensures the Gaussian field will retain its statistical properties if there are many modes to average over.  The distribution of magnitudes could simply be $P(\deltahat) d\deltahat = \dirac(\deltahat - \sqrt{VP(k)}) d\deltahat$ for example.  Now as $L$ is reduced to a finite value again, the value of $\deltahat(\boldk)$ in an infinitesimally small bin in $k$-space, consisting of an average of power from many modes in the $L \to \infty$ case, will be drawn from the Rayleigh distribution.}$^,$\footnote{Some authors mistakenly state that the magnitude should be drawn from a Gaussian distribution.}
\beq
P_{\rm R} (\deltahat) d \deltahat = \frac{\deltahat}{\sigma^2} e^{-\frac{\deltahat^2}{2\sigma^2}} d\deltahat
\eeq
where the variance $\sigma^2 = VP(k)/2$ ensures that $\langle | \deltahat(k) |^2 \rangle = VP(k)$. 
This is equivalent to drawing both the real and imaginary parts of $\hat \delta(\boldk)$ from Gaussian distributions with variance also 
$\sigma^2 = VP(k)/2$ \citep{bagla_padmanabhan_2004, klypin_holtzman_1997}. 

The density field is real, so $\deltahat(\boldk)$ must satisfy the usual Hermitian constraints.  Particular attention must be paid to the eight modes at the corners of the $\deltahat$ array which are forced to be real (those modes where each Cartesian component of $\boldk$ is either 0 or $k_{\rm Ny}$).  The value of $\deltahat$ at these $\boldk$ should be drawn from a (real) Gaussian while still satisfying $\langle \deltahat(\boldk) \rangle$ = 0 and $\langle |\deltahat(\boldk)|^2 \rangle = V P(\boldk)$.  This issue is especially important for the $\boldk = 0$ mode (if $P(0) \ne 0$, as we will consider in \S\ref{s:convolving_pk}) because there are no other modes with similar amplitudes of $k$ to dilute any error.

Obviously $\boldk$
cannot go to infinity in a computer program, so a Nyquist cutoff is introduced
which can be of order the natural Nyquist frequency $\pi/\Delta x$ given by the
particles.  The field $\deltahat(\boldk)$ can then be
inverse Fourier transformed into the real-space density field, defined
on a Eulerian grid.  To convert the Eulerian density into a Lagrangian
representation of particles, the Zeldovich approximation is widely used \citep{zeldovich_1970,edwf_1985}:
\begin{manyeqns}
\boldx &=& \boldq + \boldPsi(\boldq) \\
\dot \boldx &=& \dot \boldPsi(\boldq) \\
-\nabla \cdot \boldPsi &=& \delta(\boldq) \label{e:psi_delta}
\end{manyeqns}
where $\boldq$ is the initial position of the particle, and $\boldx$ is the final position.  Equation~\ref{e:psi_delta} is meant to apply at any epoch, and since $\delta$ scales with time as the growth factor (equation~\ref{e:growth_fxn}), $\boldPsi(\boldq) = \dbar(a) \boldPsi_0(\boldq)$ if $\boldPsi_0$ is the linear gravitational field at the present 
epoch.\footnote{A simple check of the Zeldovich appoximation is offered by the approximation
$\deltahat(\boldk) = \frac{V}{N}\sum \exp{(-i\boldk \cdot [\boldq + \boldPsi(\boldq)])} 
\approx \frac{V}{N} \sum (-i\boldk \cdot \boldPsi(\boldq)) \exp{(-i \boldk \cdot \boldq)}$ so that $\deltahat(\boldk) = -i \boldk \cdot \hat \boldPsi(\boldk)$ for small $\boldk \cdot \hat \boldPsi$.  This is just the Fourier transform of equation~\ref{e:psi_delta}.}$^,$\footnote{
Most authors neglect to mention that the growth function should be normalized to its present-day value, an acceptable oversight only when $D(1) = 1$, as in Einstein-de Sitter cosmologies.}
The velocity equation becomes
\beq
\dot \boldx = \frac{\dot D(a)}{D(1)} \boldPsi_0(\boldq)
\eeq
with $\dot D(a)$ given by equation~\ref{e:growth_fxn_dot}.  These are comoving velocities; multiply by $a$ to get the normal physical 
velocities.\footnote{Also mind the convention that cosmological distances are usually expressed in units of $\mpch$ but velocity units usually do not contain the factor of $h$.}

In practice, it is convenient to use Fourier techniques to perform differentiation: e.g., $\partial/\partial x_i \to i k_i$, $\nabla^2 \to - k^2$.  The usual technique is to use Poisson's equation in Fourier space to convert $\deltahat$ to a potential, and then to convert the real-space potential to the gravitational field $\boldPsi$ via finite-differencing
\citep{edwf_1985, bagla_padmanabhan_2004, power_etal_2003}.
However, if one has sufficient computer memory, I see no reason not to convert $\deltahat$ to the gravitational field directly via
\beq
\hat \boldPsi(\boldk) = \frac{i \boldk}{k^2} \deltahat(\boldk)
\label{e:greens}
\eeq
because the density field in the box is a \emph{continuous} field with discrete particles, in contrast to a discrete field such as a Eulerian grid.  Similarly, the discrete forms of the Green's functions are sometimes used \citep{edwf_1985, bagla_padmanabhan_2004}, but again, they are not necessarily relevant to Lagrangian cosmological simulations. 

\section{The failure of the conventional method}
\label{s:sampling_pk}

Sampling the power spectrum (or the field $\deltahat(\boldk)$) leads to a gross underestimate of the real space variance (e.g., $\sigma_8$), the correlation function, and nonlinear effects.  It is well known that the two-point correlation function and power spectrum are Fourier transform pairs.  But sampling the power spectrum is equivalent to multiplying it by a $\shah$ function of spacing $\Delta k = 2\pi/L$.  By the convolution theorem, this sampling is equivalent to convolving the continuous correlation function with a $\shah$ of spacing $L$; this is a simply a manifestation of aliasing, except in real space instead of the more usual $k$-space.  As \citet{pen_1997} pointed out, the statistical properties of real space, such as the correlation function and mass variance in spheres, can be grossly misrepresented. 

I have used cmbfast \citep{seljak_zaldarriaga_1996} to generate a power spectrum at the present epoch with the cosmological parameters $\Omegam = 0.27, \Omegal = 0.73, \Omega_{\rm b} = \Omegam - \Omega_{\rm CDM} = 0.044, H_0 = 71\ {\rm km}\,{\rm s}^{-1}\,{\rm Mpc}^{-1}$ and spectral index $n_{\rm s} = 0.97$, normalized to $\sigma_8 = 0.84$.

\begin{figure}
\includegraphics[width=3.5in]{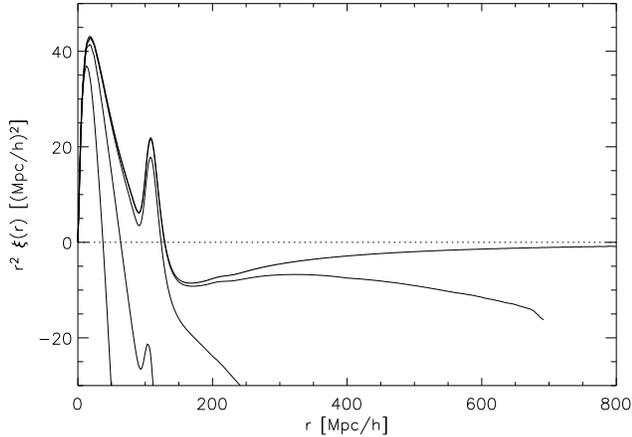}
\caption{Aliased correlation functions for $L = (100, 200, 400, 800) \mpch$ (thin lines).  These are obtained by sampling the power spectrum on a grid, inverse Fourier transforming, and collapsing angular dependence.  Because the 3-D correlation function array is cubical, information about $\xi(r)$ can be obtained for $r \in (L/2, \frac{\sqrt{3}}{2}L)$ from the ``corners'' of the $\xi(\boldx)$ array.  This information about $\xi(r)$ can be thought of as arising from correlations between the density field in opposite corners of the $\delta(\boldx)$ box.  The grid sizes were $128^3, 128^3, 256^3,$ and $512^3$ respectively, which are large enough that this plot would look little different in the limit of infinite grid size.  The thick line is the continuous inverse Fourier transform of the cmbfast power spectrum.}
\label{f:xi_wrong}
\end{figure}

Figure~\ref{f:xi_wrong} shows correlation functions obtained by sampling this power spectrum with spacing $\Delta k$, performing an inverse fast Fourier transform to get a discrete 3-dimensional correlation function, and then averaging values of $\xi(r)$ in the grid for each unique $r$.  These aliased correlation functions are representative of the correlation function one would measure in a density field after generating initial conditions with the conventional method of \S\ref{s:conventional_ic} (neglecting linear growth).  Also shown is the infinite-volume correlation function, obtained by an exact Fourier transform of the power spectrum (see equation~\ref{e:fourier_isotropized}); the baryon bump is visible at $r \approx 120 \mpch$.  It is easy to see that box sizes as small as 50 or $100 \mpch$ have profound effects on the correlation function at virtually all length scales.  Note that since $P(0) = 0$, the integral of the correlation function over the box must equal zero; this explains the fact that the zero-crossing $r_0$ of the correlation function occurs at substantially smaller $r$ than in the continuous case, especially for box sizes $L \lesssim 2 r_0$.

This problem is frequently stated as the finite value of $L$ cutting off perturbations with wavelengths greater than $L$ \citep[cf.][]{gelb_bertschinger_1994b, baugh_gaztanaga_efstathiou_1995, tormen_bertschinger_1996, knebe_dominguez_2003, bagla_ray_2004}.  Actually, the finite value of $L$ cuts out all modes $\boldk$ that do not lie on the lattice $\shah$ of spacing $\Delta k$.  However, the effect of sampling may be more pronounced at super-$L$ modes because the three-dimensional power spectrum has a cusp at $\boldk = 0$ (i.e., $P(\boldk) \propto k^n$ for small $k$ where $n \approx 1$ for currently favored near scale-invariant cosmologies).

A preliminary understanding of the effect of super-$L$ modes on the mass variance in spheres can be obtained by integrating
\beq
\sigma_8^2 = \frac{1}{2\pi^2} \int_{2\pi/L}^\infty P(k) |\hat W(k)|^2 k^2 dk
\label{e:s8_with_lower_bound}
\eeq
for various values of $L$, where $\hat W(k)$ is given by equation~\ref{e:wk}. \citet{gelb_bertschinger_1994b} found that the underestimate of $\sigma_8$ for $L = 50 \mpch$ was about 10\% for an Einstein-de Sitter cosmology.  Since currently favored $\Lambda$CDM cosmologies have more power on larger scales (for a given power spectrum normalization), this underestimate would be even more severe if calculated with the cmbfast power spectrum used in this paper.

However, the integral of equation~\ref{e:s8_with_lower_bound} is not quite the correct method for determining the effect of super-$L$ modes.  Since the real-space box is a periodic finite volume, the Fourier convention of equation~\ref{e:fourier} should be used.  Then the equation for $\sigma_8$ is accurately modified to
\beq
\sigma_8^2 = \frac{1}{V} \sum_{\boldk \in (i,j,k)\Delta k} P(\boldk) | \hat W(\boldk) |^2
\label{e:s8_discrete}
\eeq
and the contribution to $\sigma_8^2$ from super-$L$ modes is given by $P(0)/V$, scaled by $\dbar^2(a)$ if earlier epochs are considered.  
I will return to this point in \S\ref{s:s8_result}, but first I must discuss how to get the value of the power spectrum at $\boldk=0$ by convolving the power spectrum.

\section{Convolving the power spectrum} \label{s:convolving_pk}

From the point of view of the \citet{press_schechter_1974} and \citet{hklm_1991} theories of nonlinear structure formation, \citet{pen_1997} argued that the \emph{real space} statistical properties should be accurately modeled inside the simulation box.  This is conceptually done by multiplying the real-space correlation function by the window function representing the box geometry; i.e. setting $\xi(x_i) = 0$ for $|x_i| > L/2$ where $x_i$ is a Cartesian coordinate.  From the convolution theorem, this is equivalent to convolving the power spectrum with the Fourier transform of the window function.  Finally, the real-space box is made periodic by convolving it with a $\shah$ of spacing $L$, which is equivalent to sampling the power spectrum with a $\shah$ of spacing $\Delta k$.  It is the crucial \emph{convolution of the power spectrum before sampling} that ensures that the real space statistical properties inside the box will be correct.  The mass variance in spheres of radius $R$ will be accurately modeled for box sizes $L > 4R$.

I found that it was easiest to convolve the power spectrum by first inverse Fourier transforming the continuous power spectrum into the correlation function, truncating the correlation function at $L/2$, and then Fourier transforming back.  The relevant isotropized Fourier transform equations are:
\begin{manyeqns}
\label{e:fourier_isotropized}
\xi(r) &=& \frac{1}{2\pi^2} \int_0^\infty P(k) \frac{\sin{kr}}{kr} k^2 dk \\
P_L(k) &=& 4\pi \int_0^{L/2} \xi(r) \frac{\sin{kr}}{kr} r^2 dr
\end{manyeqns}
Upon sampling this \emph{convolved power spectrum} $P_L(\boldk)$, the correlation function of the resulting real-space field will be exact in a sphere of radius $L/2$, and the mass variance of spheres will be exact for radii $R < L/4$.  It is conceptually possible to ensure that the correlation function is exact in the entire cubical box instead of just its inscribed sphere, but this would be numerically more challenging, and it is unclear that it would be of any benefit because of the anisotropy of the box on large scales.

The sampling of the \emph{convolved} power spectrum at a given $\boldk$ can be understood as taking a suitable weighted average of the power in a bin of characteristic size $\Delta k^3$ centered on $\boldk$ \citep{pen_1997}.  This will have a particularly pronounced effect on the $\boldk=0$ mode, due to the cuspy nature of the power spectrum at $\boldk = 0$.  Another way of interpreting the sampling of a convolved power spectrum is that it is the Fourier equivalent of convolving the truncated correlation function with a $\shah$ of spacing $L$, i.e., making it periodic (and introducing aliasing error in real space if the correlation function was not truncated!).

\begin{figure*}
\includegraphics[width=\textwidth]{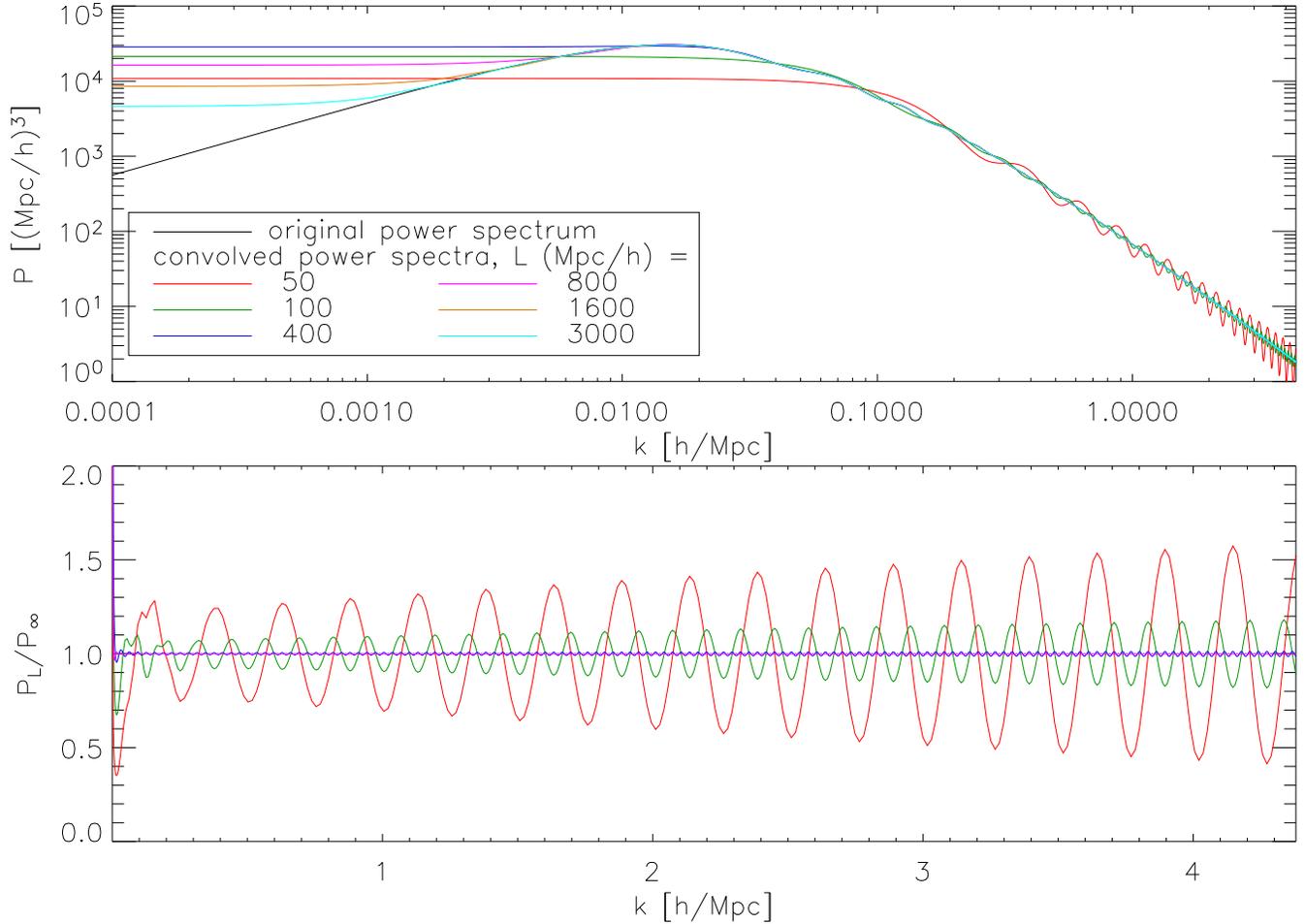}
\caption{The cmbfast power spectrum (black line) and convolved power spectra for $L = (50, 100, 400, 800, 1600, 3000) \mpch$ (colored lines).  The bottom panel shows the convolved power spectra normalized by the cmbfast power spectrum on a linear scale.  (The 1600 and $3000 \mpch$ cases are not plotted in the bottom panel.)  The ringing in the convolved power spectra has wavelength $4\pi/L$.  Note that each convolved power spectrum is intended to be sampled with a specific interval $\Delta k = 2\pi/L$ (in three dimensions), but I have plotted them as continuous functions for illustration.  }
\label{f:convolved_pk}
\end{figure*}

\begin{figure}
\includegraphics[width=3.5in]{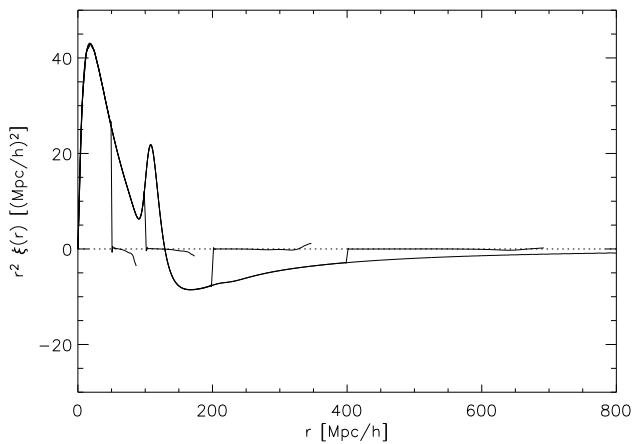}
\caption{Same as figure~\ref{f:xi_wrong}, except that the appropriate convolved power spectra are sampled for $L = (100, 200, 400, 800) \mpch$.  As intended, the correlation function matches the $L \to \infty$ limit for $r < L/2$, and is 0 for $r > L/2$.}
\label{f:xi_correct}
\end{figure}

Convolved power spectra for various values of $L$ are presented in figure~\ref{f:convolved_pk}.  The original power spectrum is again given by cmbfast.  The resulting correlation functions for these power spectra are shown in figure~\ref{f:xi_correct}, which can be compared directly to figure~\ref{f:xi_wrong}.  As expected the correlation functions are exact within $L/2$.

\begin{figure}
\includegraphics[width=3.5in]{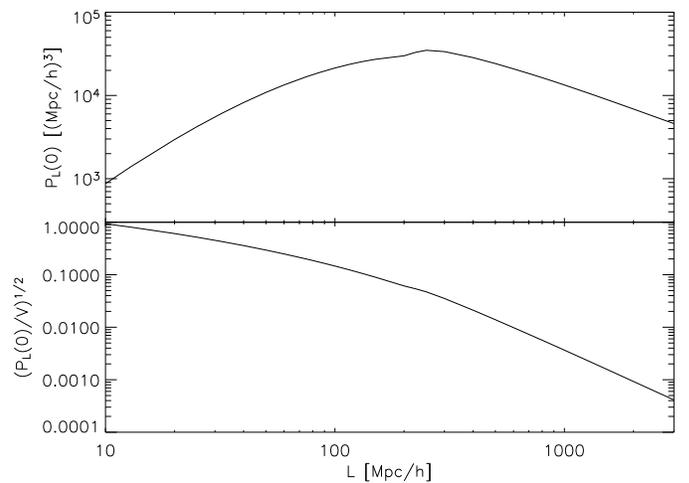}
\caption{Power in the DC mode vs.\ box size $L$ (top panel) and its dimensionless version, the characteristic dimensionless overdensity corresponding to $P_L(0)$ (bottom panel).  I used the cmbfast power spectrum normalized to $\sigma_8 = 0.84$.}
\label{f:dcmode}
\end{figure}

\section{The DC mode}\label{s:dc_mode}
In order to get the real-space statistics of the density field correct within the box, it is clear that the power spectrum must be convolved as described in \S\ref{s:convolving_pk} before sampling.  The generation of initial conditions can then proceed as usual, as in \S\ref{s:conventional_ic}, with one additional nontrivial correction: $P_L(0)$ is no longer assumed to be zero.  Figure~\ref{f:dcmode} shows what this power is for various box sizes; ideally one would run a simulation with $L \to \infty$ and $P_L(0) \to 0$.\footnote{I am assuming Newtonian gravity even for the largest boxes, as is usually done \citep[e.g.,][]{evrard_etal_2002}.}
The relevant Green's function (equation~\ref{e:greens}) becomes undefined at $\boldk = 0$.  What should one do about this DC mode?

Note that the amplitude of the DC mode is drawn from a Gaussian, so it will be necessary to run an \emph{ensemble} of simulations with different random number seeds.  

But first, consider a single realization.  The DC mode corresponds to the imposition of a constant overdensity at present epoch $\Delta_0$ drawn from a Gaussian distribution of variance $P_L(0)/V$.  Since it is a large-scale mode, its amplitude will evolve according to the linear growth function (equation~\ref{e:growth_fxn}): $\Delta = \dbar \Delta_0$.  For small $\Delta$, one could evolve an $N$-body simulation ignoring the DC mode, and then add the contribution from $\Delta$ to an evolved simulation by violating mass conservation and scaling the mass of each particle by $1+\Delta$.  
Alternatively, one could rescale the box of an evolved simulation, effectively pushing the particles as in the mode-adding procedure of \citet{tormen_bertschinger_1996}.  \citet{cole_1997} showed that, in addition, one should rescale cosmic time in order to more accurately emulate long-wavelength mode coupling.
This coupling of long-wavelength modes (and by extension, the DC mode) to all other modes can be thought of as follows.  A large-scale overdensity effectively modifies the local value of $\Omegam$ (and $\Omegal$) in its domain of the universe.  That is, structures in overdense regions grow faster.  I discuss this quantitatively below and in Appendix~\ref{s:unibox_mapping}.

It would be even more accurate to incorporate the effects of DC mode coupling into the simulation itself.  Since $\Delta > 0$ corresponds to a box which expands slower than the $\Delta = 0$ case, and vice versa for $\Delta < 0$, we just need to determine $\abox(t)$, where $t$ is cosmic time, so that the modified equation of motion, $\ddot \boldx + 2\dot \boldx  \dot a_{\rm box} / \abox= -\nabla \phi/ \abox^2$, accurately encapsulates the effects of DC mode coupling.
I will use the subscripts ``box'' and ``uni'' to refer to any parameter that relates to a simulation with $\Delta_0 \ne 0$, and $\Delta_0 = 0$, respectively (i.e., $\auni$ is the scale factor of the universe, $\Omega_{\rm m,uni}$ is given by the cosmological model but $\Omega_{\rm m, box}$ may differ, etc.).  

Since $\auni(t)$ is known from equation~\ref{e:hubble}, it is sufficient at first to derive the relationship between $\abox$ and $\auni$.  From a Eulerian viewpoint, writing the ratio of the densities of the box and the universe results in $\abox/\auni = (1 + \Delta)^{-1/3}$.  From a Lagrangian viewpoint, $\abox/\auni = (L/2 - \Psi)/(L/2)$ where $\Psi$ is a small displacement equal to the force field by the Zeldovich approximation: $\nabla \cdot \boldPsi = \Delta$.  Considering a test particle at coordinates $(L/2, 0, 0)$ where the origin coincides with the center of the box, $\Psi = L\Delta/6$.  Therefore,
\beq
\frac{\abox}{\auni} = 1 - \frac{1}{3}\Delta.
\label{e:abox_auni}
\eeq
I will use this relationship between $\abox$ and $\auni$ on the grounds that Lagrangian perturbation theory is, generally speaking, more accurate than Eulerian perturbation theory \citep{bouchet_etal_1995}.  Obviously both approaches agree as $\Delta \to 0$.

Now $\abox(t)$ is known, so the $N$-body code could be modified to reflect this change in the equation of motion.  However, one could avoid rewriting $N$-body code (and gain insight into the physics of the DC mode) by modifiying the cosmological parameters of a given realization with $\Delta \ne 0$ as follows:
\begin{manyeqns}
\label{e:unibox_mapping}
\hobox &=& \houni \frac{1}{1+\phi} \\
\ombox &=& \omuni (1+\phi)^2 \\
\olbox &=& \oluni (1+\phi)^2
\end{manyeqns}
\beq
\phi = \frac{5}{6}\frac{\Omegam}{D(1)} \Delta_0.
\label{e:phi}
\eeq
I provide the derivation in Appendix~\ref{s:unibox_mapping}.
This is the prescription for running an $N$-body simulation with a nonzero DC mode. Modify the cosmological parameters as appropriate for a given realization, run the simulation as normal, but map the cosmic time according to equation~\ref{e:abox_auni}.  That is, if you are interested in the simulation at epoch $z_{\rm uni}$, run the simulation until $z_{\rm box}$ given by $1+z_{\rm box} = (1+z_{\rm uni})/(1 - \dbar\Delta_0/3)$.

In their numerical simulations, \citet{frenk_etal_1988} used essentially the same scheme as outlined here to treat the DC mode.  They ran sets of simulations with the DC mode equal to $\{0, \pm 1\}$ times the rms overdensity, i.e.~$\sqrt{P_L(0)/V}$, instead of drawing the DC mode from a Gaussian as in this paper.  They also synchronized their ensemble at $z = 6$ with every realization having the same box size ($2\ {\rm Mpc}$), thus varying the integrated mass in the box.
In contrast, the present approach synchronizes the box size at $z \to \infty$ (see Appendix~\ref{s:unibox_mapping}) and uses the same box-integrated mass among the realizations.

\section{Ensembles of simulations}\label{s:ensembles}
To test the new method for generating initial conditions, I have run several ensembles of simulations.  I will refer to initial conditions generated in the traditional way, by sampling the power spectrum, as $P$-sampled IC's.  Initial conditions generated with the new method of convolving the power spectrum before sampling can be called $\xi$-sampled IC's because the correlation function will be accurately reproduced in the box.\footnote{The correlation function is not actually ``sampled'' in the usual sense.}

An \emph{ensemble} of simulations is a set of $N$-body simulations with the same cosmological parameters and technical parameters (e.g., $N, V$), but with different random number seeds.  
For simulations running on $\xi$-sampled IC's, the effective cosmological parameters of a given realization within the ensemble change as described in \S\ref{s:dc_mode}, but the realization still represents a part of the universe that the ensemble is intended to model. 

For each ensemble, I generated initial conditions to second order in Lagrangian perturbation theory \citep[2LPT,][]{scoccimarro_1998,bouchet_etal_1995} for the following output redshifts: 
\bea
\log{[1/(1+z_{\rm uni})]} =  -3.0, -2.8, -2.6, \nonumber \\
 -2.4, -2.2, -2.0, -1.8, -1.6.
\eea
I used the last of these outputs as initial conditions to the $N$-body code, tree particle mesh \citep[TPM,][]{bode_ostriker_2003}.  The output redshifts from TPM were
\bea
1/(1+z_{\rm uni}) = 10^{-1.4}, 10^{-1.2}, \nonumber \\
0.10, 0.15, 0.20, 0.25, \cdots 1.00
\eea

All the ensembles presented herein comprised 100 realizations of the cosmology $\Omegam = 0.27, \Omegal = 0.73, h = 0.71$.
The only parameters varied in this work are $L$, $N$, and whether I used $P$- or $\xi$-sampled IC's.  I used the same 100 random number seeds for all ensembles, and because of the way the Fourier modes are populated in the initial conditions, the same large-scale structures, scaling as $L$, develop in all ensembles for a given random seed.

Note a subtle change in terminology at this point.  Instead of \emph{inputting} parameters like $\sigma_8, P(k)$ into initial conditions, I will now present and analyze \emph{measured} values for $\sigma_8$, $P(k)$ and $\xi(r)$ from the particle distributions resulting from the combination of 2LPT and TPM.

\section{The mass variance in $8 \mpch$ spheres $\sigma_8$} \label{s:sigma8}
\subsection{Method}
To calculate $\sigma_8$ for a given epoch of a single realization, it is possible to use equation~\ref{e:s8_discrete} if the power spectrum is known.  The discrete Fourier transform, for which the well-known fast Fourier transform is an implementation, introduces aliasing error into the power spectrum.  Even if the continuous Fourier transform is used (equation~\ref{e:fourier1}), it is impossible to have $k \to \infty$ in a computer program.  It seems better to be able to control the errors easily by using a real-space Monte Carlo integration scheme:
\bea
\sigma_8^2 &=& \frac{1}{N_{\rm r}} \sum^{N_{\rm r}} (\delta(\boldx_{\rm r}) * W)^2 \label{e:s8_mc} \\
\delta(\boldx_{\rm r}) * W &=& \frac{V_{\rm uni}}{N} \frac{N_{\rm sphere}}{V_{\rm sphere}} - 1
\eea
where $N_{\rm r}$ random points $\boldx_{\rm r}$ within the box volume are chosen to serve as integration abscissae.   $V_{\rm uni}$ is the volume of the box \emph{if $\Delta_0$ had been zero}.  The window function $W$ is a spherical tophat of comoving radius $8 \mpch$ and volume $V_{\rm sphere}$; the number of particles within that sphere (at location $\boldx_{\rm r}$) is simply $N_{\rm sphere}$.  The error on $\sigma_8^2$ can then be estimated as
\beq
\Delta \sigma_8^2 = \sqrt{ \frac{ \langle (\delta * W)^4 \rangle - \langle (\delta * W)^2 \rangle^2 }{N_{\rm r}} }
\eeq
\citep[\S7.6]{press_etal_1992} where brackets indicate an average over $N_{\rm r}$ random points.  Unless otherwise noted, all values of $\sigma_8^2$ in this paper have been calculated to 2\% accuracy (for individual realizations), which corresponds to approximately 1\% accuracy on $\sigma_8$ itself.

Because $\sigma_8^2$ is a volume average (assuming ergodicity), the value of $\sigma_8^2$ for the ensemble is a weighted average of the values of $\sigma_8^2$ from each realization, and the weight of each realization is $\abox^3$.  

\begin{figure}
\includegraphics[width=3.5in]{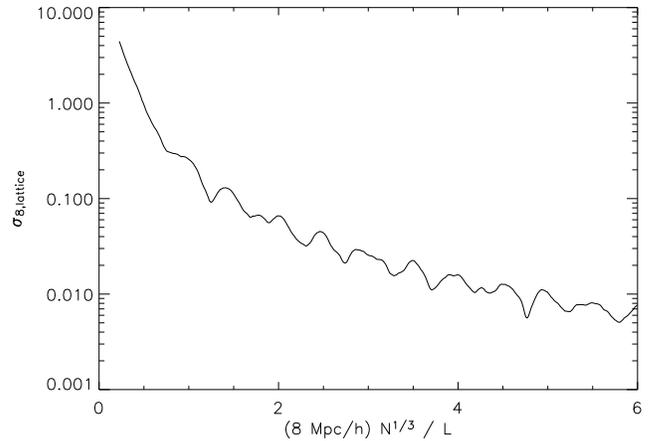}
\caption{The value of $\sigma_{8,{\rm lattice}}$ as a function of $R/\Delta x$, where the mean interparticle separation $\Delta x = L/N^{1/3}$ and the radius of the spherical tophat is $R = 8 \mpch$.  Values are accurate to 1\%.}
\label{f:s8_lattice}
\end{figure}

Let $\sigma_{8,{\rm lattice}}$ be the value of $\sigma_8$ from an unperturbed lattice of particles, and let $\sigma_{8,{\rm Poisson}}$ be the expected value of $\sigma_8$ from a Poisson distribution of particles.  These two quantities will be important to the discussion in \S\ref{s:s8_result} below.

It is easiest to calculate $\sigma_{8,{\rm lattice}}$ with a Monte Carlo integrator, using equation~\ref{e:s8_mc}.  Figure~\ref{f:s8_lattice} shows the result, which only depends on the ratio of the radius of the spherical tophat $R = 8 \mpch$ to the spacing between particles $\Delta x = L/N^{1/3}$.  Naturally, as $\Delta x$ decreases, the effects of particle discreteness become less severe, and $\sigma_{8,{\rm lattice}}$ decreases.  The wiggles in figure~\ref{f:s8_lattice} can be thought of as interactions between the spherical symmetry of a sphere and the translational symmetry of a lattice as the relative scales change.

A Poisson distribution of particles results from placing particles randomly throughout the box.    It is well known that a Poisson distribution has $P(\boldk) = V/N$ for all $\boldk$.   Note that $V/N$ is the expectation value, and power spectra of realizations and finite ensembles will fluctuate about this value.  The expectation value of $\sigma_{8,{\rm Poisson}}^2$ should properly be calculated from equation~\ref{e:s8_discrete} because the real-space box is a periodic finite volume.  However, we can avoid evaluating the infinite sum by using the following trick, based on calculating $\sigma_8$ in real space.  Imagine a box of size $L > 4R$ into which we will throw down particles, randomly placed.  As we throw down particles, imagine convolving the particles with the $R = 8 \mpch$ spherical tophat so that each particle becomes a diffuse sphere.  Since the box is periodic we can assume the first particle lands in the exact center of the box without loss of generality.  The second particle may land so that the distance between the two particles is either less than $2R$, or greater than $2R$.  In the former case, the ``domains of influence'' of the two spheres overlap and it is exactly this ``interaction'' that our real space $\sigma_8$ calculator must identify.  In the latter case, the distance between the two particles is irrelevant.  In other words, $L$ could have been twice as large and the distance between the two particles could have been twice as large, and our $\sigma_8$ calculator would not have been able to tell the difference.  (The introduction of more particles just generalizes to a superposition of many two-particle ``interactions.'')  Therefore, as long as $L > 4R$, the size of the box does not matter for our $\sigma_8$ calculator.  If $L < 4R$, then the second particle could ``interact'' with the first particle on both sides due to periodicity, and this double-interaction would not be a Poisson process.  The above argument implies that the infinite sum of equation~\ref{e:s8_discrete} must yield the same (exact) answer as the continuous integral
\beq
\sigma_8^2 = \frac{1}{2\pi^2} \int_{0}^\infty P(k) |\hat W(k)|^2 k^2 dk
\label{e:s8_continuous}
\eeq
for a Poisson distribution in a box of size $L > 4R$, where $\hat W(k)$ is given by equation~\ref{e:wk}.  The value of the integral is $3V/4\pi R^3 N$, so $\sigma_{8,{\rm Poisson}} = 0.1193$ for $L = 100 \mpch$, $N = 32^3$. 
Strictly, for a Poisson distribution the number of particles in the box $N$ fluctuates on the order of $\sqrt{N}$, but I will use the term ``Poisson'' even when $N$ is fixed.  The effect of using exactly $N$ particles is to set the DC mode to zero, and the missing contribution to $\sigma_8^2$ from this single mode is $1/N$, too small to fret about in this work.

\begin{figure*}
\includegraphics[width=\textwidth]{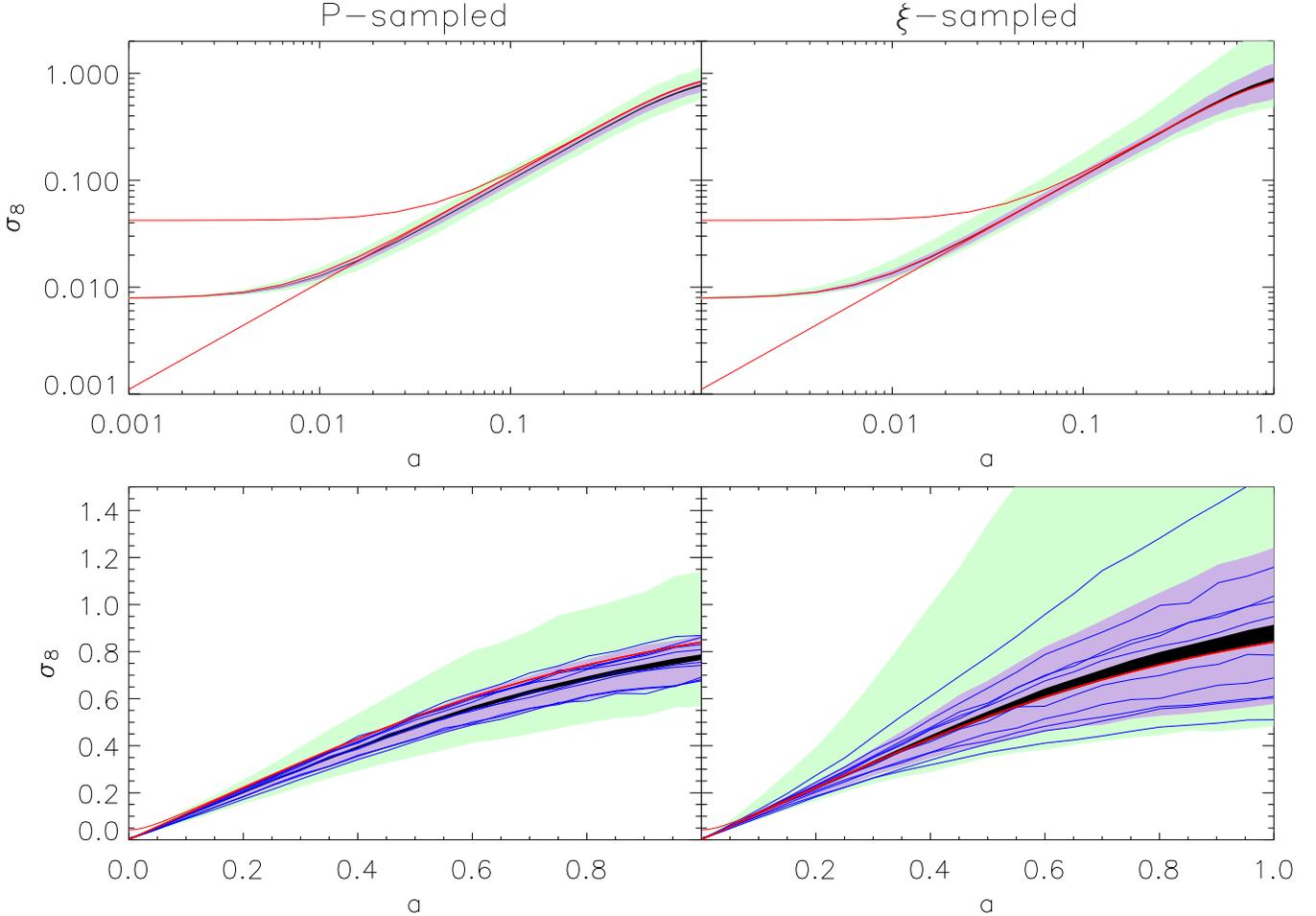}
\caption{$\sigma_8$ as a function of scale factor $\auni$ for $P$-sampled IC's (left panels) and $\xi$-sampled IC's (right panels).  The ensembles had $L = 50 \mpch$, $N = 32^3$.  For a given $\auni$, I calculated $\sigma_8$ with a Monte Carlo integrator for each realization.  The light green region shows the range of $\sigma_8$ for the 100 realizations.  The purple region shows the 68.3\% confidence region on individual realizations, and the black ``line'' shows the 68.3\% confidence region for the ensemble average.  The linear theory expectation of $\sigma_8$'s evolution is plotted as the bottom red line. The middle red line is the theoretical value with the expected lattice contribution added in quadrature; the top red line is the theoretical value with the appropriate Poisson contribution added in quadrature.
The bottom panels plot the same information as the top panels, except using a linear scale, which makes clearer the discrepancy between the theoretical and calculated evolution of $\sigma_8$ in the $P$-sampled case, and the agreement in the $\xi$-sampled case.  In the bottom panels I have also plotted, with blue lines, ten of the 100 realizations to show qualitatively individual errors due to the Monte Carlo integrator.  I chose the same ten random seeds for both the $P$- and $\xi$-sampled cases.  These individual evolution tracks also show qualitatively that a realization with a small value of $\sigma_8$ early in the simulation has a small value later in the simulation as well.}
\label{f:s8_example}
\end{figure*}

\subsection{Results} \label{s:s8_result}
Figure~\ref{f:s8_example} shows $\sigma_8$ for two ensembles with $L = 50 \mpch$, $N = 32^3$ differing in whether they use $P$- or $\xi$-sampled IC's.  Since a plot with all 100 realizations would be too confusing, I have plotted the range of $\sigma_8$ for the 100 realizations in light green.  The 15th and 84th percentiles are bounded by the purple region, which approximately indicates the 68.3\% confidence region for individual realizations.  I used a bootstrap method to determine the 68.3\% confidence region on the actual ensemble average, plotted as the black region (more optimistically, the thick line).  The ensemble average can be compared to the theoretical expectation $\sigma_{8,{\rm theory}} = \dbar(a) \times 0.84$, plotted as the lower red line.  It is evident that a systematic bias exists in the value of $\sigma_8$ for the ensemble in the $P$-sampled case; it is too low.  The $\xi$-sampled case corrects this bias as expected.  I will discuss this effect more quantitatively below.

The behavior of the ensemble value of $\sigma_8$ at early times is well approximated by $( \sigma_{8,{\rm lattice}}^2 + \sigma_{8,{\rm theory}}^2 )^{1/2}$, plotted as the middle red line in figure~\ref{f:s8_example}.  Why does the contribution from the lattice add in quadrature?  Recall that the power spectrum of the lattice of particles is zero everywhere except for modes with each component an integer multiple of $2k_{\rm Ny}$ (excluding $\boldk = 0$).  Applying a small displacement field to the particles results in a power spectrum for the particles that accurately models the input power spectrum for sub-Nyquist wavenumbers.  If the displacement field is small, then the lattice part of the power spectrum will be barely modified for $k \sim 2 k_{\rm Ny}$, but will be more greatly modified for very large $k$.  From equation~\ref{e:s8_discrete}, the measured value of $\sigma_8^2$ can be decomposed into two sums.  The first sum, over all sub-Nyquist wavenumbers, measures $\sigma_{8,{\rm theory}}^2$ from the information in the displacement process.  The second sum, over all super-Nyquist wavenumbers, approximately measures $\sigma_{8,{\rm lattice}}^2$ because the weight of $|\hat W(\boldk)|^2$ is greatest for those modes which are least affected by the displacement, i.e., the modes $k \sim 2 k_{\rm Ny}$.

The upper red line in figure~\ref{f:s8_example} shows $( \sigma_{8,{\rm Poisson}}^2 + \sigma_{8,{\rm theory}}^2 )^{1/2}$.  It is clear that Poisson noise is not a good model for the underlying discretization of matter into mass particles for most of the logarithmic range of the simulation.  In fact, vestiges of the initial lattice can be seen in these simulations to $z \sim 2$ (and even later times for simulations with larger $\Delta x$), so there is no reason to model the discretization of matter into particles as a Poisson distribution before these epochs.

\begin{figure*}
\includegraphics[width=\textwidth]{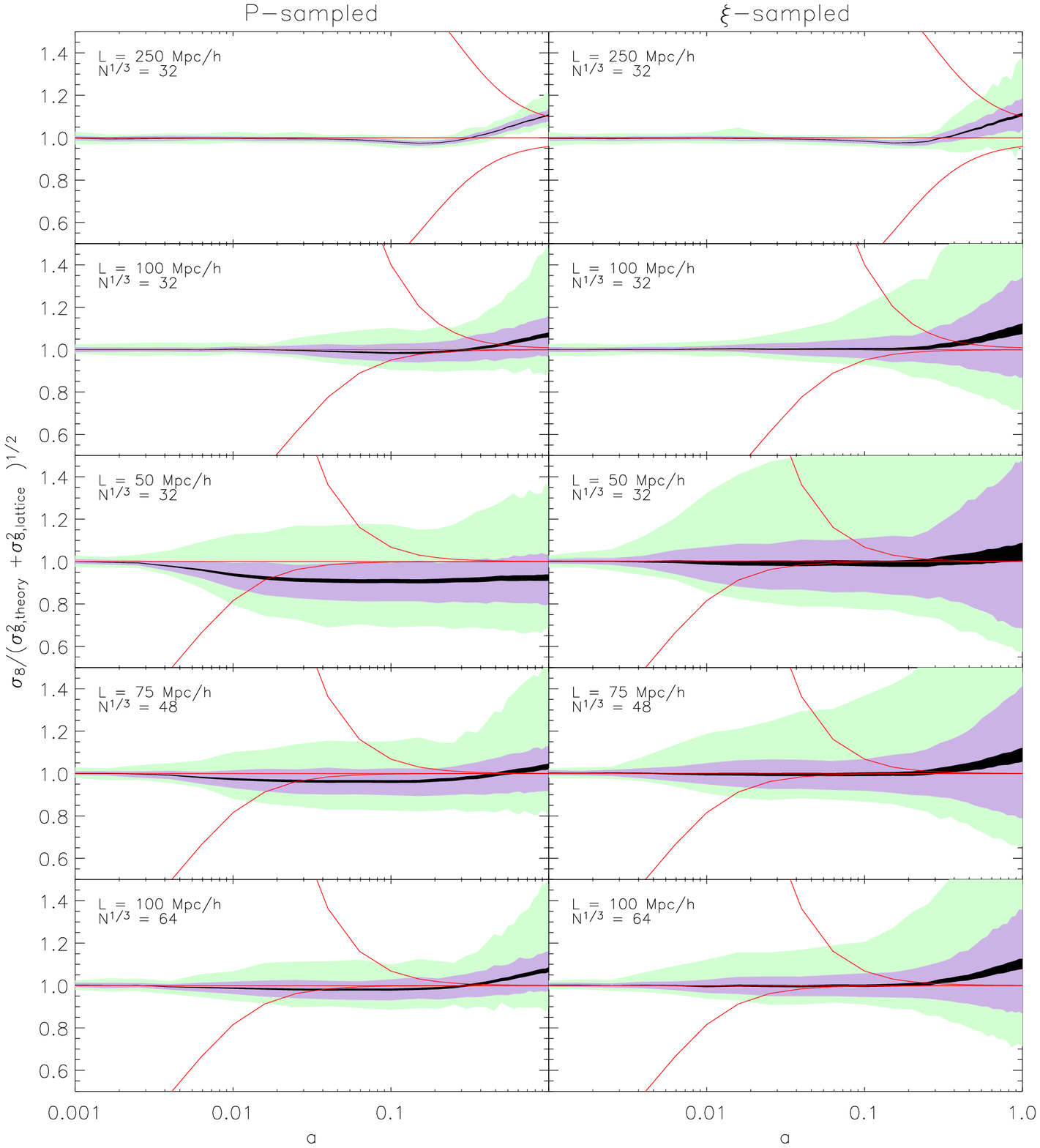}
\caption{Evolution of $\sigma_8$ of ten ensembles, normalized to the linear theory prediction plus the expected lattice contribution $(\sigma_{8,{\rm theory}}^2 + \sigma_{8,{\rm lattice}}^2)^{1/2}$.  The light green, purple, and black regions, and the three red lines are as in figure~\ref{f:s8_example}.}
\label{f:s8_all}
\end{figure*}

Figure~\ref{f:s8_all} shows the evolution of $\sigma_8$ for both $P$-sampled and $\xi$-sampled IC's for five different combinations of technical parameters $(L, N)$.  In this figure all curves are normalized to $(\sigma_{8,\rm{theory}}^2 + \sigma_{8,{\rm lattice}}^2)^{1/2}$, so to the extent that the simulations evolve according to linear theory the black region should match the lower red line $\sigma_{8,{\rm theory}}/(\sigma_{8,\rm{theory}}^2 + \sigma_{8,{\rm lattice}}^2)^{1/2}$.  The middle two panels, representing the case $L = 50 \mpch$, $N=32^3$, show the same information as figure~\ref{f:s8_example} for comparison.  The two rows above it have $N = 32^3$ as well, but $L = (100, 250) \mpch$ so that $\sigma_{8,\rm{lattice}}$ is large.  The bottom two rows of the figure show the effect of increasing $L$ while keeping $\Delta x$ fixed.  As expected, in the $P$-sampled case, the discrepancy between $\sigma_8$ and $\sigma_{8,{\rm theory}}$ (for intermediate times) is reduced for larger $L$, and completely eliminated in the $\xi$-sampled case.  It also appears that nonlinear evolution of $\sigma_8$ at late times is enhanced using the larger box size.  This is presumably due to the greater density of modes available for coupling, and cannot be solved using $\xi$-sampled IC's with small box sizes.  However, it is fair to claim that using $\xi$-sampled IC's solves the first order problem of the discrepancy between $\sigma_8$ of the ensemble and $\sigma_{8,{\rm theory}}$.

\begin{figure*}
\includegraphics[width=\textwidth]{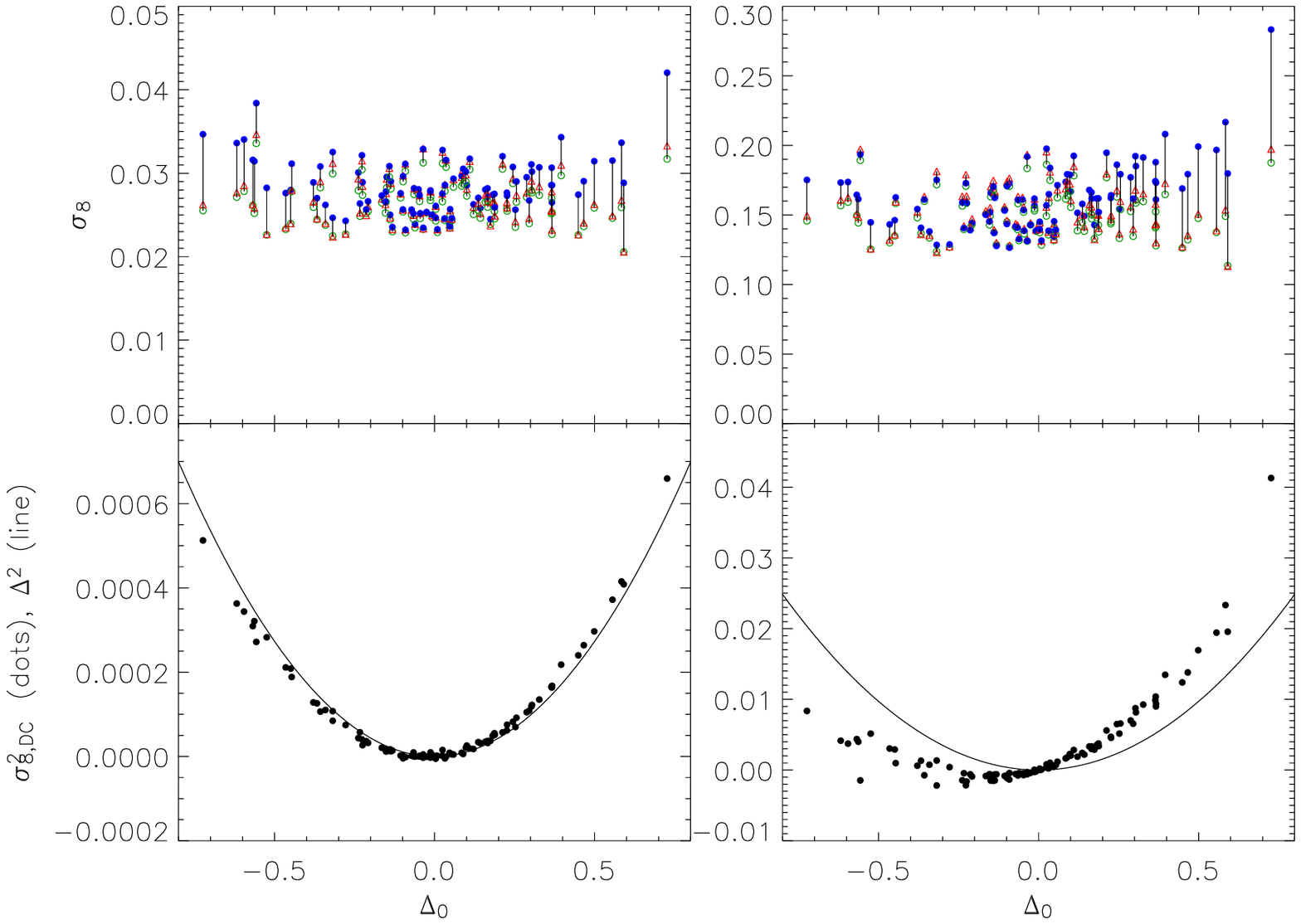}
\caption{The upper panels show 100 individual realizations' values of $\sigma_8$ for $L = 50 \mpch$, $N = 32^3$ for $z = 38.81$ (left panels) and $z = 5.67$ (right panels) with particle positions generated with the second order Lagrangian perturbation theory initial conditions code.  The green open circles are for $P$-sampled IC's; the blue filled circles are for $\xi$-sampled IC's, and the red triangles are for $\xi$-sampled IC's with the DC mode artificially set to zero.  All three cases used the same 100 random number seeds.  A  thin black line connects the range of values for each realization.  The values of $\Delta_0$ from the $\xi$-sampled case serve as the abscissae.  Values of $\sigma_8$ are accurate to 0.25\%.  The lower panels plot the 100 realizations' values of $\sigma_{8,{\rm DC}}^2 = \sigma_{8,\xi-{\rm sampled}}^2 - \sigma_{8,P-{\rm sampled}}^2$.  The expected value of $\Delta^2 = (\dbar(a) \Delta_0)^2$ is also shown.}
\label{f:s8_v_delta}
\end{figure*}

Let's now take a step back to examine the effect of $P$-sampled vs.\ $\xi$-sampled IC's on individual realizations.  Figure~\ref{f:s8_v_delta} shows the values of $\sigma_8$ for the 100 realizations of $L = 50 \mpch$, $N = 32^3$ at $z = 38.81$ (left panels) and $z = 5.67$ (right panels).  For this figure I just used the 2LPT initial conditions code for both of these redshifts, and I calculated values of $\sigma_8^2$ to 0.5\% accuracy.  In the upper panels, note that the values of $\sigma_8$ for the $\xi$-sampled case (blue filled circles) tend to be greater than those for the $P$-sampled case (green open circles), and that the discrepancy is greater for larger $|\Delta_0|$ (of the $\xi$-sampled IC's).  I have also generated particle positions using the convolved power spectrum but artificially setting the DC mode to zero; the values of $\sigma_8$ for these configurations are plotted as red triangles.  By comparing these three types of IC's, it is apparent that most of the discrepancy between the $P$- and $\xi$-sampled IC's comes from the DC mode itself.  

The lower panels in figure~\ref{f:s8_v_delta} show the difference in $\sigma_8^2$ between the $\xi$-sampled and $P$-sampled IC's, $\sigma_{8,{\rm DC}}^2 = \sigma_{8,\xi-{\rm sampled}}^2 - \sigma_{8,P-{\rm sampled}}^2$.  If, as I claimed above, the dominant difference between $\xi$- and $P$-sampled IC's manifests itself in the DC mode, $\sigma_{8,{\rm DC}}^2$ is approximately the contribution to $\sigma_8^2$ from the DC mode.  From equation~\ref{e:s8_discrete}, the expected contribution from the DC mode is just $\Delta^2 = \dbar^2(a) P_L(0)/V = (\dbar(a)\Delta_0)^2$, plotted as a line.  The expected contribution agrees well with $\sigma_{8,{\rm DC}}^2$, especially for higher redshift when, presumably, mode coupling has less effect.

Now we can answer the question: what is the effect of using $\xi$-sampled IC's instead of $P$-sampled IC's on the \emph{ensemble} value of $\sigma_8$?  For early times, the weighting of each realization (by $\abox^3$ as properly required) is not important, so the ensemble contribution $\sigma_{8,{\rm DC,e}}^2$ is an integral over the realization values $\sigma_{8,{\rm DC}}^2$:
\bea
&& \sigma_{8,{\rm DC,e}}^2 = \nonumber \\
&& \frac{1}{\sqrt{2\pi P_L(0)/V}} \int_{-\infty}^\infty \sigma_{8,{\rm DC}}^2\, e^{-\frac{\Delta_0^2}{2 P_L(0)/V}} d\Delta_0
\eea
where $P_L(0)/V$ is the variance of the DC mode given by the convolved power spectrum, and $\Delta_0$ is the dimensionless DC mode of a single realization (at present epoch).  Since $\sigma_{8,{\rm DC}}^2 \approx (\dbar(a) \Delta_0)^2$, the expected contribution to the ensemble average $\sigma_{8,{\rm DC,e}}^2 = P_L(0)/V$ at the present epoch.  The error in $\sigma_8$ due to the DC mode when using $P$-sampled IC's is approximately $\frac{1}{2} (P_L(0)/V)/\sigma_{8,{\rm theory}}^2$, where $P_L(0)/V$ can be read off of figure~\ref{f:dcmode}.  Figure~\ref{f:dcmode} uses a normalization of $\sigma_{8,{\rm theory}} = 0.84$ (which cancels out in the estimate of this error).  The error is about 6\% for $L = 50 \mpch$, and reduces to 1\% for $L = 120 \mpch$.  This is not to say that a $P$-sampled simulation of the latter size is accurate to 1\% in all respects; note that the characteristic dimensionless overdensity for $L = 120 \mpch$ is about 12\% (for $L = 50\mpch$, 30\%).

\section{The correlation function $\xi(r)$} \label{s:xi}
\subsection{Method}
The well-known formula for the correlation function is
\beq
\xi(\boldx) = \langle \delta(\boldx') \delta(\boldx + \boldx') \rangle
\eeq
where the brackets indicate an ensemble average, or space average assuming ergodicity.  But first, we will need the correlation function of a single realization.  Substitute $\delta = \rho/\rho_0 - 1$, where $\rho_0$ is the ensemble (universe) average, and isotropize to obtain the formula for the correlation function of particles in a box of size $L_{\rm box}$:
\bea
\xi(r) = \left(\frac{\auni}{\abox}\right)^3 \frac{L_{\rm uni}^3}{N} \frac{\langle N_{\rm shell} \rangle}{V_{\rm shell}} \nonumber \\
- 2\left(\frac{\auni}{\abox} \right)^3 + 1.
\label{e:corrfxn}
\eea
Here $N_{\rm shell}$ is the number of particles in a thin spherical shell of radius $r$ and volume $V_{\rm shell}$ surrounding a single particle, and $\langle N_{\rm shell} \rangle$ is an average over all the particles in the box (or a random subset).  

For $L/2 < r < \sqrt{3}L/2$, the ``shell'' used in equation~\ref{e:corrfxn} becomes the intersection of a spherical shell with the box.  In this way, it is possible to obtain information about the correlation function from particles on opposite corners of the box, but it must be stressed that these are the scales on which the box is anisotropic so it is unlikely that the correlation function on these scales should ever be used in rigorous analysis.

The correlation function of the ensemble is a simple weighted average of the correlation functions of the individual realizations, and the weight of each realization is $\abox^3$.

\begin{figure*}
\includegraphics[width=\textwidth]{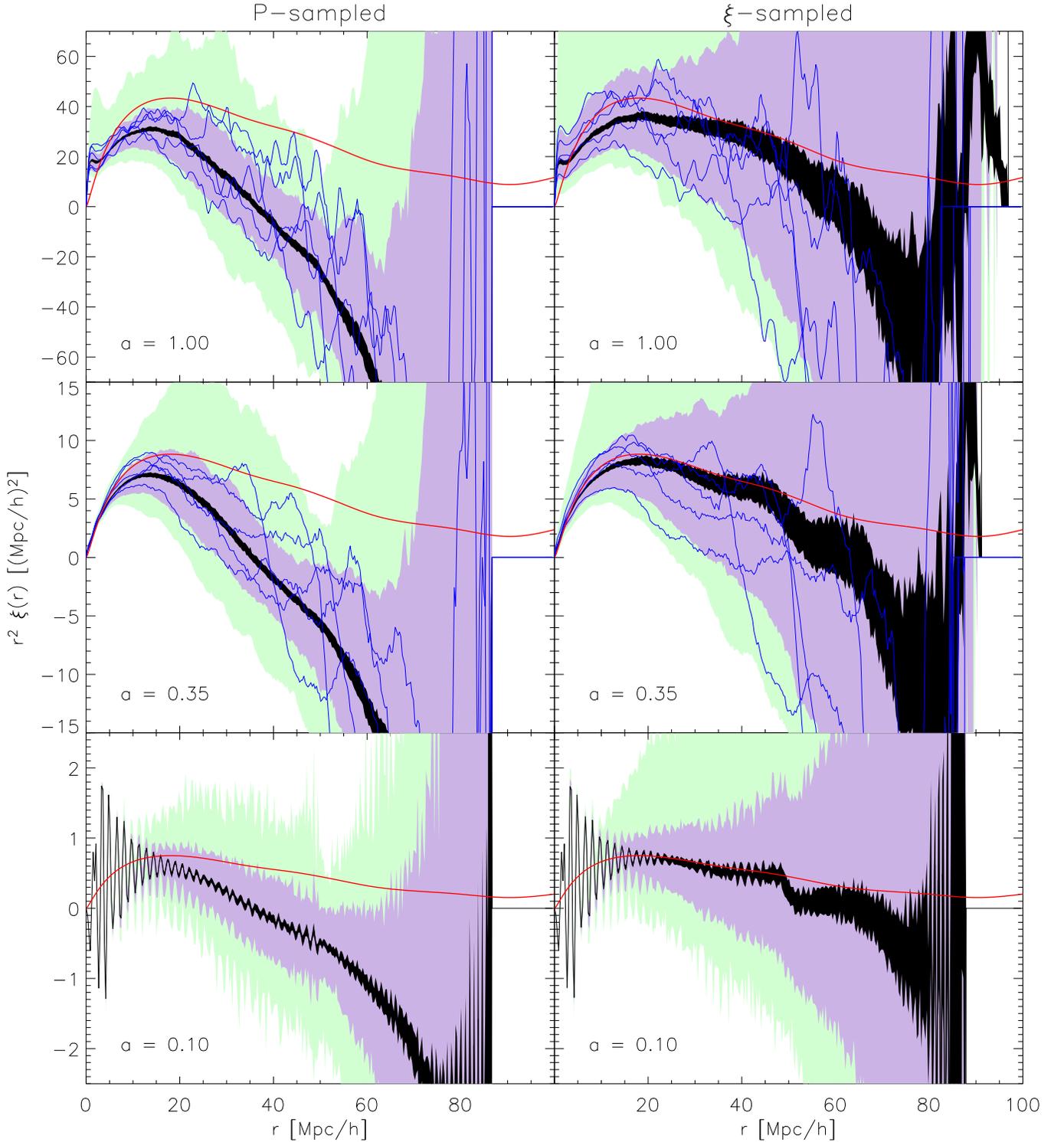}
\caption{Correlation functions for the $L=100 \mpch$, $N=64^3$ ensembles with $P$-sampled IC's (left panels) and $\xi$-sampled IC's (right panels) for three output $\auni$, labeled.  The black region is the 68.3\% confidence region on the ensemble correlation function; the purple region is the approximate 68.3\% confidence region on individual realization correlation functions, and the light green region bounds all 100 realizations' correlation functions.  In the upper four panels, five individual correlation functions for example are overplotted in blue (the random number seeds chosen match across each panel).  The red line is the linear theory prediction.  As in figures~\ref{f:xi_wrong} and~\ref{f:xi_correct}, information about the correlation function is available for $r > (\abox/\auni) L/2$ from the corners of the simulation cube, but it is unlikely that this information should ever be used in rigorous analysis.}
\label{f:xi_ensemble}
\end{figure*}

\subsection{Results} \label{s:xi_result}
Ensemble correlation functions are presented in figure~\ref{f:xi_ensemble} for several output redshifts for two ensembles, differing in whether they used $P$-sampled or $\xi$-sampled IC's.  These ensembles had $L = 100 \mpch$ and $N = 64^3$.  I binned the correlation function into $0.3 \mpch$ bins.  Note the large systematic offset between the measured correlation function and linear theory prediction in the $P$-sampled case, which is largely corrected in the $\xi$-sampled case for $r < L/2$, as intended.  The ``ringing'' in the correlation function at early times is due to the initial lattice and is erased cleanly as the simulation evolves.

\section{The power spectrum $P(k)$}\label{s:pk}
\subsection{Method}
To eliminate aliasing error, I used the brute force Fourier transform (equation~\ref{e:fourier1}) to determine the power spectrum of each realization.  I chose the following abscissae ($\boldk = (i,j,k)\Delta k$): 
\bea
(i,j,k) \in \pm \{ 0, 1, 2, 3, \cdots 16, 18, 20, \cdots \nonumber \\
32, 36, 40 \cdots 64, 72, 80, \cdots 128 \}
\label{e:pk_abscissae}
\eea
to achieve a balance between good coverage in $k$-space and computational feasibility.  
It is important to consider biases introduced by this sampling of abscissae.  For example, the power spectrum of a lattice of $64^3$ particles has a large spike at  $\boldk = (64,64,64) \Delta k$ and is zero in the bins surrounding that $\boldk$, so the power in that mode should not be taken to be representative of the power in the surrounding bins.  

For $\xi$-sampled IC's, the effective box size of each realization is modified, and therefore $\Delta k$ is different among the realizations in an ensemble.  The power spectrum of the ensemble, like the correlation function and $\sigma_8^2$, can be determined from the realizations' power spectra by weighting each by $\abox^3$ before averaging.  (This can be seen from the linearity of the Fourier transform of the correlation function.)  For the $\xi$-sampled case, the abscissae of each realization shift, so I binned the power in $k$-space before averaging.  I used a binning corresponding to the spacing of $k$ in equation~\ref{e:pk_abscissae}.  To be consistent I binned the $P$-sampled case correspondingly.  It was necessary to keep careful track of the number of modes at each $k = |\boldk|$ to do the averaging correctly.

\begin{figure*}
\includegraphics[width=\textwidth]{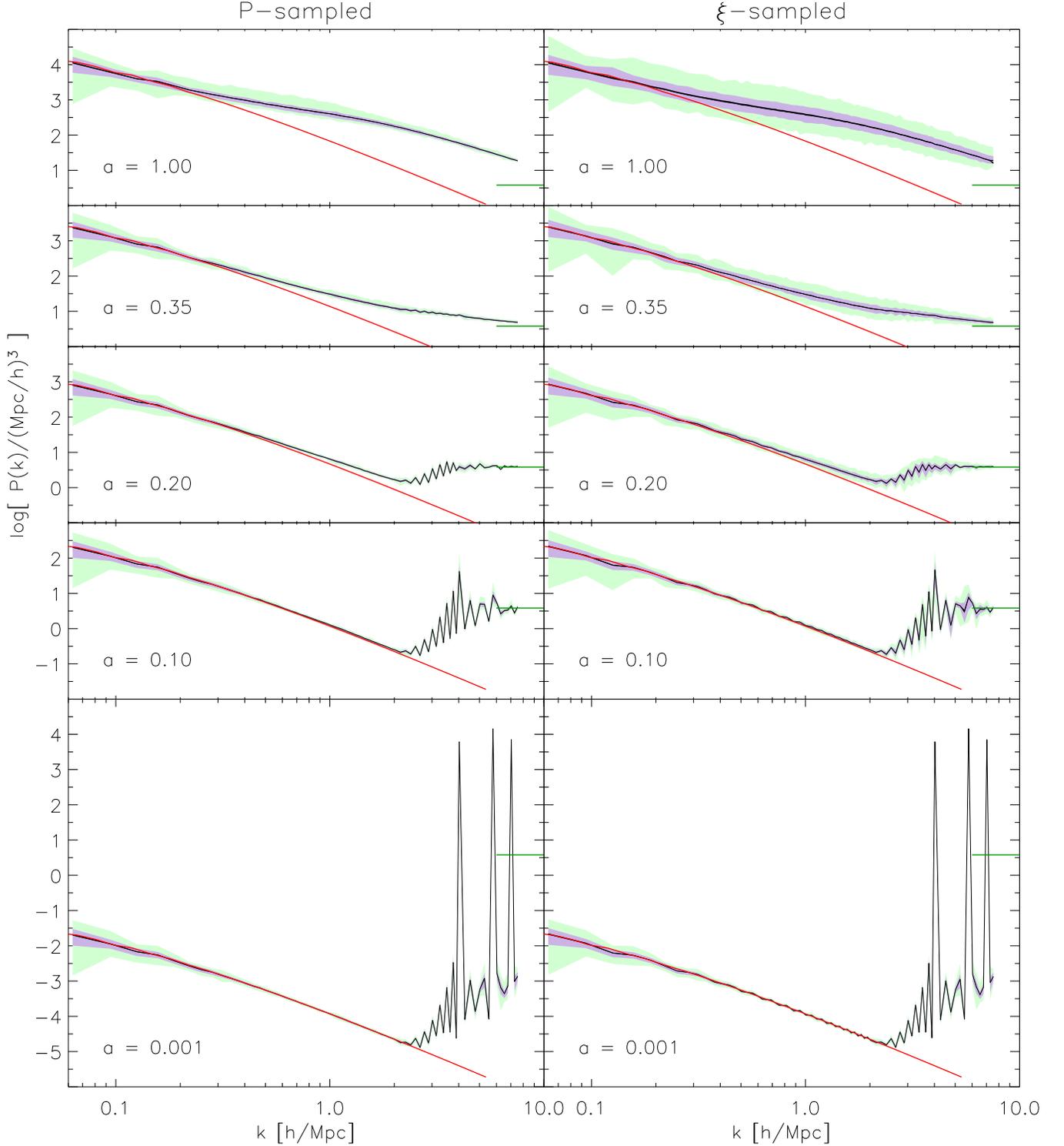}
\caption{Power spectra for the $L=100 \mpch$, $N=64^3$ ensembles with $P$-sampled IC's (left panels) and $\xi$-sampled IC's (right panels) for five output $\auni$, labeled.  The black line, purple region, light green region, and red line have the same meanings as in figure~\ref{f:xi_ensemble}.  On the right side of each panel is a green line marking the expected Poisson noise contribution to the power spectrum, $V/N$.
}
\label{f:pk_ensemble}
\end{figure*}

\subsection{Results}
Figure~\ref{f:pk_ensemble} shows power spectra for the same two ensembles discussed above in \S\ref{s:xi_result}, for various epochs.  The DC mode is omitted from these plots.  Not much is different between the $P$-sampled and $\xi$-sampled cases, except for the oscillations in the $\xi$-sampled case for early epochs which are subsequently washed out, and the much greater dispersion in normalization for the $\xi$-sampled case at late epochs.  The latter effect is a consequence of the mode coupling of the DC mode to all other modes; a box with a large-scale overdensity effectively emulates a universe with a larger $\Omegam$, etc.  However, as we have seen directly in figure~\ref{f:s8_v_delta} and indirectly in figure~\ref{f:xi_ensemble}, the DC mode is the most important effect of using $\xi$-sampled IC's.  Therefore the difference in ensemble power spectra (besides the DC mode) between $P$- and $\xi$-sampled IC's is not as dramatic as the difference in correlation functions seen in figure~\ref{f:xi_ensemble}.

\section{Subtracting Poisson noise}\label{s:poisson}
When measuring power spectra of $N$-body simulations, the Poisson noise contribution is sometimes subtracted to arrive at a more ``true'' power spectrum that represents the underlying continuous density field \citep[e.g.,][]{smith_etal_2003, jing_2004}.  The rationale is that the discreteness of particles introduces a delta function in the correlation function, or that a Poisson distribution is representative of a field with ``no'' clustering.  However, $N$-body simulations start with a pre-initial lattice (or glass) which is not a Poisson distribution of particles.  As shown in figures~\ref{f:s8_example} and~\ref{f:s8_all}, the Poisson contribution to $\sigma_8$ is a bad model until \emph{at least} rather late epochs.  Similarly, \citet{baugh_efstathiou_1994} and \citet{baugh_gaztanaga_efstathiou_1995} pointed out that the Poisson contribution cannot be subtracted until the power spectrum evolves above the level $V/N$.  Looking at the $\sigma_8$ results of figures~\ref{f:s8_example} and~\ref{f:s8_all}, it is unclear whether or not the Poisson contribution is a bad model at very late epochs, though, since the lattice is mostly erased by that time.  

Another insight is provided by the power spectra of figure~\ref{f:pk_ensemble}.  
Note the spikes in the power spectrum at early epochs.  These are caused by the lattice of particles and initially have power $V$ at $\boldk = (i,j,k) 2\pi/\Delta x$, $\boldk \ne 0$.  These spikes can be called Bragg peaks \citep{gabrielli_etal_2005} by analogy with x-ray scattering of crystals.  Particle evolution smooths out these Bragg peaks, so that at $a \sim 0.2$ in figure~\ref{f:pk_ensemble} (before the imposed perturbation spectrum has significant effect on high $k$) the power becomes Poissonian at high $k$.  That is, the power has redistributed to an expectation value $V/N$ over the $N$ bins in $k$-space surrounding the initial Bragg peak.   The integrated power has been neither magnified nor suppressed because the power at these high $k$ is due to discreteness, not evolution of a collisionless fluid.  The crucial point is that there is no initial power in the lattice at $\boldk = 0$, so at sub-Nyquist $k$, the resulting Poisson power spectrum contribution \emph{does not exist, and should not be subtracted}.  The lack of the Poisson power spectrum at sub-Nyquist scales can be seen from the middle panels of figure~\ref{f:pk_ensemble}, where there is a dip in the power spectrum below the Poisson level at $k \sim 2.5\, h\,{\rm Mpc}^{-1}$.  However, it remains unclear if the discreteness power at high $k$ can migrate inwards to affect all $k$ by the end of the simulation, and therefore if some reasonable model of the discreteness contribution to the power spectrum should be subtracted.  The problem of the spreading of initial discreteness power to the wavenumbers of interest in the simulation will be left as an open question.  The issue may be made moot merely by using negligible values of $V/N$ in simulations of the future.

\section{Summary}\label{s:summary}
I have described and implemented $\xi$-sampled IC's, as suggested by \citet{pen_1997}.  The advantage of $\xi$-sampled IC's is that they accurately reproduce the real-space statistical properties of the universe within the box.  $P$-sampled IC's, on the other hand, systematically underestimate the mass variance in spheres and the shape of the correlation function.  Most analytic nonlinear theories of structure formation \citep[e.g.,][]{press_schechter_1974, hklm_1991} are best understood in terms of real-space gravitational collapse, so it is important to get the real-space statistical properties correct in $N$-body simulations.  Another way of making this point is to imagine extracting a finite size box from an underlying infinite density field (not worrying about periodicity, which comes naturally from sampling the power spectrum).  The mean density in randomly placed boxes will not exactly equal the cosmological mean density, and this sample variance should be included in $N$-body simulations.  
As implied by \S\ref{s:s8_result}, the most important contribution of $\xi$-sampled IC's is to incorporate the DC mode, so one could use $P$-sampled IC's but treat the DC mode specially using the techniques of \S\ref{s:dc_mode}, as done by \citet{frenk_etal_1988}.  However, to ensure that the real-space statistical properties are accurately encoded within the box, one should use the convolved power spectrum as described in \S\ref{s:convolving_pk}.

Thus there are two distinct but related effects of using $\xi$-sampled IC's.  To reiterate, the first is that the \emph{mean} real-space statistical properties of an ensemble become unbiased.  For example, for $L = 50 \mpch$, $\xi$-sampled IC's correct the 6\% underestimate in $\sigma_8$ of $P$-sampled IC's.  
$\xi$-sampled IC's also fix the shape of the correlation function to match the input correlation function for $r < L/2$.  
The second effect is that the \emph{variance} of DC modes becomes nonzero.  For example, the characteristic overdensity of $50 \mpch$ boxes is 30\%!  While the tests between the two methods in this paper have been limited to certain statistics, one could imagine that this latter effect would be \emph{extremely} important in constructing halo catalogs from an ensemble of simulations, for example.

One could, of course, reduce the problem of inaccurately modeled real-space statistics with $P$-sampled IC's by using very large boxes.  But since each doubling of the box size increases computation time by about an order of magnitude, one could ask if it would be better to run one $L = 100 \mpch$ simulation or ten realizations of a $L = 50 \mpch$ ensemble.  The answer is not obvious and probably depends on the problem to be solved.  The former simulation necessarily ignores the DC mode variance, so $\sigma_8$ will be biased.  The latter ensemble will have an unbiased $\sigma_8$ but will be unable to reproduce the correlation function between $50$ and $100 \mpch$.

I have also argued that the commonly assumed Poisson noise model for the discreteness of particles is not a trivially added contribution to the power spectrum.  More investigation about the relationship, in the power spectrum, between the discreteness effects and the underlying continuous perturbations is warranted. 

Code to generate initial conditions to second order in Lagrangian perturbation theory \citep{scoccimarro_1998} is available at 
\verb+http://www.astro.princeton.edu/~esirko/ic+.

\acknowledgments
I thank David Spergel, Ue-Li Pen, and Francesco Sylos Labini for helpful discussions.
This work is partially supported by NASA HQ Award \#NNG04GK55G.
The computations in this paper used facilities provided in part by 
NSF grant AST-0216105.

\appendix
\section{Derivation of the ``uni'' $\to$ ``box'' mapping} \label{s:unibox_mapping}
This section outlines the derivation of the ``uni'' $\to$ ``box'' mapping of cosmological parameters in a $\Delta \ne 0$ realization (equations~\ref{e:unibox_mapping} and~\ref{e:phi}).

At early times, $\abox = \auni$ and $\Omegam \to 1$, so simplifying equation~\ref{e:hubble} results in
\beq
\frac{\dot{a}_{\rm box}}{\abox} = \hobox \sqrt{\frac{\ombox}{\abox^3}}, \qquad
\frac{\dot{a}_{\rm uni}}{\auni} = \houni \sqrt{\frac{\omuni}{\auni^3}}
\eeq
and it is easily seen that $(\hobox/\houni)^2 = \omuni / \ombox $ in order to accurately model the early epochs.  Therefore let $\hobox = \houni/(1+\phi), \ombox = \omuni (1+\phi)^2$, and $\olbox = \oluni (1+\gamma)^2$ where $\phi$ and $\gamma$ are parametrizations of the modifications of the cosmological parameters.

Taking the time derivative of equation~\ref{e:abox_auni}, 
\beq
\frac{\dot{a}_{\rm box}}{\abox} = \frac{\dot{a}_{\rm auni}}{\auni} -\frac{ \dot \Delta/3}{1 - \Delta/3}.
\eeq
Inserting equations~\ref{e:hubble},~\ref{e:growth_fxn_dot}, and~\ref{e:abox_auni}, squaring both sides, substituting the $\phi$ and $\gamma$ parametrizations, and then expanding to second order in $\phi$, $\gamma$, and $\Delta$ results in:
\bea
\lefteqn{ \frac{\Omegam}{a^3}\left(1 + \Delta + \frac{2}{3}\Delta^2\right) +
\Omegal (1+2\gamma - 2\phi + \gamma^2 - 4\gamma\phi + 3\phi^2) + } \nonumber \\
& & \frac{1}{a^2} \left(1 - 2\phi + \frac{2}{3}\Delta + 3\phi^2 - \frac{4}{3}\phi\Delta + \frac{1}{3}\Delta^2 \right) - \nonumber \\
& & \frac{\Omegam}{a^2} \left(1 + \frac{2}{3}\Delta + \frac{1}{3}\Delta^2 \right) -
\frac{\Omegal}{a^2} \left(1 + \frac{2}{3}\Delta + \frac{1}{3}\Delta^2 \right) = \\
& & \frac{\Omegam}{a^3} + \Omegal + \frac{\Omegak}{a^2} -
\frac{\Omegam}{3a^2} \left( \frac{5}{D} - \frac{3}{a} - \frac{2\Omegak}{\Omegam} \right) \left(\Delta + \frac{1}{3}\Delta^2 \right) + \nonumber \\
& & \frac{\Omegam^2}{36a^2} \left( \frac{5}{D} - \frac{3}{a} - \frac{2\Omegak}{\Omegam} \right)^2 \left( \frac{\Omegam}{a^3} + \Omegal + \frac{\Omegak}{a^2} \right)^{-1} \Delta^2 \nonumber
\eea
where I have dropped the ``uni'' subscripts on the cosmological parameters and $\auni$ for notational simplicity.  Using the Eulerian form of equation~\ref{e:abox_auni} only affects some of the second-order terms.
The zeroth-order equation is trivially satisfied.  Extracting the first order equation,
\beq
\frac{\Omegam}{a^3}\Delta + \Omegal(2\gamma - 2\phi) - \frac{2}{a^2} \phi +
\frac{2 \Omegak}{3a^2} \Delta = - \frac{\Omegam}{3a^2} \left( \frac{5}{D} - \frac{3}{a} - \frac{2\Omegak}{\Omegam} \right) \Delta.
\eeq
In order for the perturbation parameters not to have a dependence on $a$, $\gamma$ must equal $\phi$.  The equation then simplifies to
\beq
\phi = \frac{5}{6}\frac{\Omegam}{D(1)} \Delta_0.
\eeq

\begin{figure}
\includegraphics[width=\textwidth]{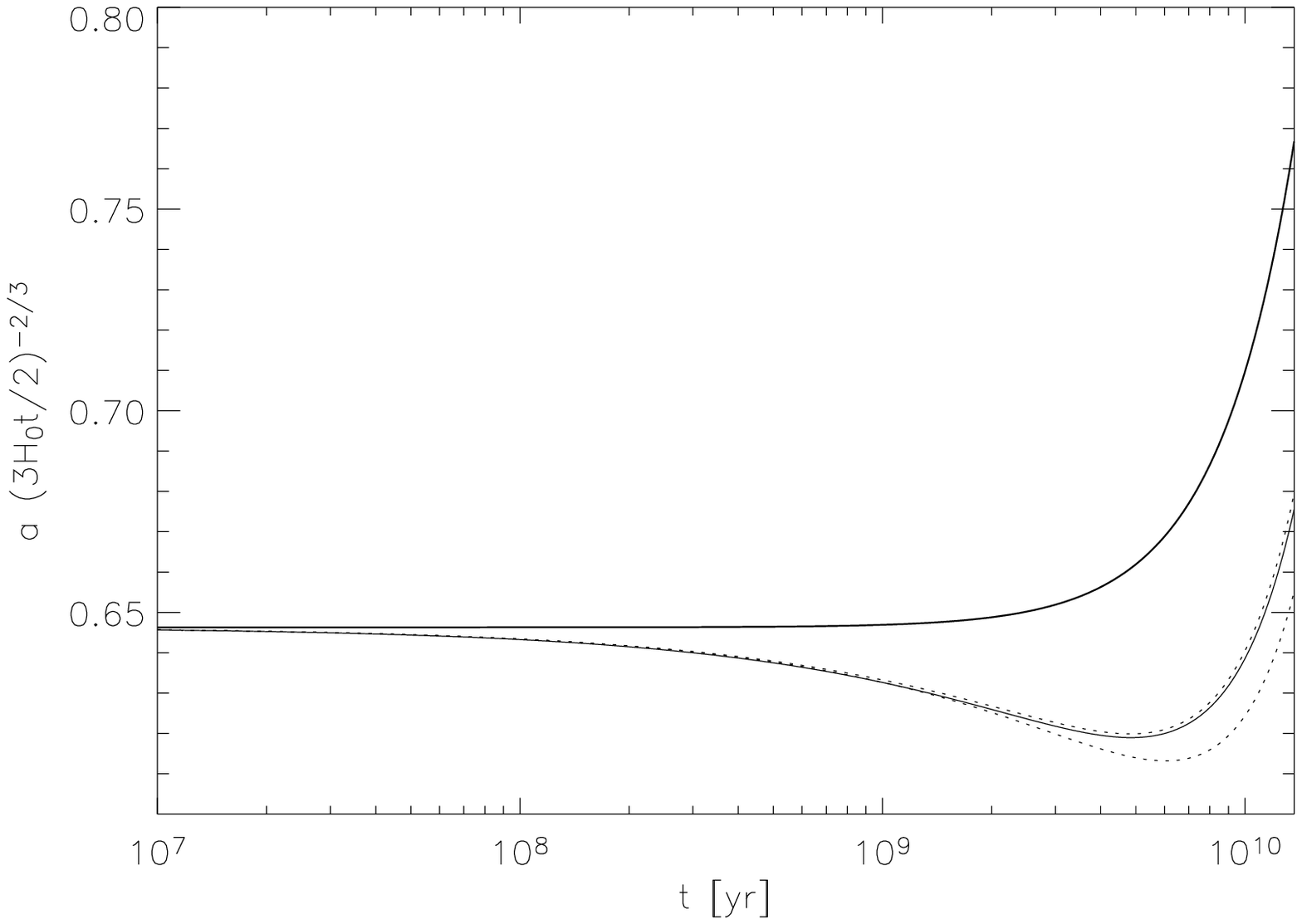}
\caption{Scale factor vs.\ cosmic time, normalized to Einstein-de Sitter expansion.  The thick line shows $\auni$ vs.\ $t$ for $\Omegam = 0.27, \Omegal = 0.73, h = 0.71$.  For this cosmology the Hubble time is $1/H_0 = 13.77 \Gyr$, and the age of the universe is $13.67 \Gyr$; the latter serves as the boundary of the plot on the right side.  For $\Delta_0 = 0.4$, the upper dotted line is the Eulerian estimate on what $\abox$ should be: $\abox = (1+\Delta)^{-1/3}\auni$.  The lower dotted line is the Lagrangian estimate $\abox = (1-\Delta/3)\auni$.  The thin line is $\abox$ vs.\ $t$ for the perturbed cosmological parameters.}
\label{f:box_v_uni}
\end{figure}

Figure~\ref{f:box_v_uni} shows the result of this approximation for $\Delta_0 = 0.4$ (quite a large value!).  The ``box'' cosmology accurately reproduces the expected behavior for $\abox(t)$.

\end{document}